\pdfoutput=1

\documentclass[12pt,a4paper,hyperref]{jhepLike}
\let\ifpdf\relax
\usepackage{ifpdf}
\usepackage{color}
\let\normalcolor\relax
\usepackage{bbm,amsmath,graphics,mathtools,cite,shuffle}
\newcommand{\fwbox}[2]{\text{\makebox[#1][c]{$\hspace{-150pt}\displaystyle#2\hspace{-150pt}$}}}
\newcommand{\fwboxL}[2]{\text{\makebox[#1][l]{$#2$}}}
\newcommand{\fwboxR}[2]{\text{\makebox[#1][r]{$#2$}}}
\newcommand{\eq}[1]{\vspace{-0.5pt}\begin{equation}\fwbox{300pt}{#1}\vspace{-0.5pt}\end{equation}}
\newcommand{\eqs}[1]{\vspace{-0.0pt}\begin{equation}\begin{split}#1\end{split}\vspace{-0.0pt}\end{equation}}
\newcommand{\fig}[3]{\raisebox{#1}{\ \includegraphics[scale=#2]{#3}}}

\newcommand{\z}[2]{(z_{#1}\!-\!z_{#2})}

\newcommand{\e}[2]{\epsilon{}_{{}_{\!}#1{}_{{}_{\!}}#2}}
\newcommand{\lek}[3]{\epsilon\hspace{-0.5pt}k_{#1,(#2 #3)}}

\newcommand{\s}[2]{s_{{}_{{}_{\!}}{}_{{}_{\!}}#1{}_{{}_{\!}}#2}}
\newcommand{\ls}[3]{s{}_{{}_{\!}#1{}_{{}_{\!}}({}_{\!}#2#3{}_{\!})}}
\newcommand{\pp}[3]{s_{{}_{{}_{\!}}{}_{{}_{\!}}#1{}_{{}_{\!}}#2{}_{{}_{\!}}#3}}
\newcommand{\ppfrac}[2]{\frac{#1}{#2}}
\newcommand{\mi}{\raisebox{0.75pt}{\scalebox{0.75}{$\,-\,$}}}
\newcommand{\pl}{\raisebox{0.75pt}{\scalebox{0.75}{$\,+\,$}}}
\renewcommand{\phi}{\varphi}
\definecolor{hblue}{rgb}{0,0,0.75}
\definecolor{hred}{rgb}{0.65,0,0.15}

\newcommand{\ek}[2]{\epsilon\hspace{-0.5pt}k_{#1 #2}}
\newcommand{\ekexpl}[2]{2\hspace{1pt}\epsilon_{#1}\hspace{-4.5pt}\cdot\hspace{-2.5pt}k_{#2}}

\preprint{2016}
\title{{\Large \mbox{Analytic Representations of Yang-Mills Amplitudes}}}
\author{{\normalsize \mbox{N.~E.~J.~Bjerrum-Bohr$^1$, Jacob~L.~Bourjaily$^1$, Poul~H.~Damgaard$^1$ and Bo~Feng\mbox{$^{2}$}}}\\
\mbox{{\mbox{$^1$}\ Niels Bohr International Academy and Discovery Center,}}\\
\mbox{{\ \;\! The Niels Bohr Institute, University of Copenhagen,}}\\
\mbox{{\ \;\! Blegdamsvej 17, DK-2100 Copenhagen \O, Denmark}}\medskip \\
\mbox{{\mbox{$^2$}\ Zhejiang Institute of Modern Physics, Zhejiang University,}}\\
\mbox{{\ \;\! Hangzhou City, 310027, People's Republic of China}}}
\date{\today}
\abstract{Scattering amplitudes in Yang-Mills theory can be represented in the formalism of Cachazo, He and Yuan (CHY) as integrals over an auxiliary projective space---fully localized on the support of the scattering equations. Because solving the scattering equations is difficult and summing over the solutions algebraically complex, a method of directly integrating the terms that appear in this representation has long been sought. We solve this important open problem by first rewriting the terms in a manifestly M\"{o}bius-invariant form and then using monodromy relations (inspired by analogy to string theory) to decompose terms into those for which combinatorial rules of integration are known. The result is the foundations of a systematic procedure to obtain analytic, covariant forms of Yang-Mills tree-amplitudes for any number of external legs and in any number of dimensions. As examples, we provide compact analytic expressions for amplitudes involving up to six gluons of arbitrary helicities.}

\begin{document}
\newpage
\section{Introduction}\label{sec:introduction}\vspace{-6pt}
One of the most fundamental quantities in theoretical particle physics is the scattering amplitude for $n$ gauge bosons. Although so essential, it is remarkable that for a long time explicit expressions for covariant $d$-dimensional scattering amplitudes of $n$ massless gauge bosons of arbitrary helicities were most easily obtained from the field theory limit of string theory (see, $e.g.$,~\cite{GSW} for a review). Conventional $d$-dimensional Feynman diagram techniques are simply way too cumbersome above a small number of external legs. The highly efficient BCFW on-shell recursion relations~\cite{Britto:2004ap,BCFW} provide a practical solution, but it would still be worthwhile to explore alternate approaches.

In the scattering equation formalism of CHY,~\cite{Cachazo:2013gna,Cachazo:2013hca,Cachazo:2013iea}, represents a completely new step towards obtaining such compact covariant expressions for amplitudes. Expressed in terms of a (reduced) $2n\!\times\! 2n$ Pfaffian, the $n$-point S-matrix element is given by a $(n\mi3)$-dimensional integral which {\it fully localizes} on the set of solutions to so-called scattering equations. A proof of the validity of this remarkable formula for any $n$ has been given in ref.~\cite{Dolan:2013isa} {and it has also been derived from the viewpoint of the ambitwistor string~\cite{Mason:2013sva,Adamo:2013tsa,Adamo:2015gia
}}. Thus, no integrations are really required to find the $n$-point covariant scattering amplitude, only a sum over solutions to a set of algebraic equations. The downside of this is that the sum scales with $n$ as $(n\mi3)!$ and finding the full set of solutions becomes difficult already at rather low values $n$. Progress has been made from a variety of different directions~\cite{Sogaard:2015dba,Dolan:2015iln,Lam:2016tlk}.\footnote{We are also aware of another approach to analytic integration---very different than what is described here---that should work for arbitrary CHY/string-theory integrands, \cite{Zlotnikov}.}

Recently, a simple set of analytic integration rules were derived. They circumvent the problem of summing over $(n\mi3)!$ solutions and provides the result of that sum based on a simple combinatorial algorithm,~\cite{Baadsgaard:2015voa,Baadsgaard:2015ifa,Baadsgaard:2015hia}. However, some of the integrals needed in order to obtain explicit expressions for covariant gauge boson amplitudes were not immediately in a form where these simple integration rules were applicable. Rather, one would first have to resort to a not entirely systematic use of integration-by-parts identities. This makes it hard to provide general and simple rules for deriving any $n$-point gauge boson scattering amplitude using this formalism.

Very recently, the issue of integration rules for more general CHY integrands has been considered from two independent directions~\cite{Gomez:2016bmv,Huang:2016zzb}. {The monodromy relations solve such problems by shifting the integration contours appropriately. That way we rewrite all integrands in terms of pieces that all have $\alpha' \to 0$ limits without further analytic continuation. Other prescriptions with less compact integrands (e.g., rewritten through also integration by parts identities) can indeed be verified to be free of such terms. However, such prescriptions appear very hard to systematize.} In this paper, we shall present a different and fully systematic solution to the problem---applicable at least to the case of integrands appearing in the CHY representation of Yang-Mills amplitudes. Interestingly, our method uses the idea of monodromy as it is applied in string theory~\cite{BjerrumBohr:2009rd,Stieberger:2009hq}. This is perhaps puzzling on two counts. First, monodromy relations in string theory  {\it a priori} only provide non-trivial relations between {\it full amplitudes}: by a sequence of contour shifts, and upon taking first real and then imaginary parts~\cite{BjerrumBohr:2009rd}, one derives KK amplitude relations~\cite{KK} and BCJ amplitude relations~\cite{BCJ}, respectively. Second, because the CHY construction is based on entirely different integrations on a set of $\delta$-function constraints, it may not seem {\it a priori} obvious why monodromy considerations can apply to that formalism. 

To understand the first issue, one should realize that monodromy in string theory is far more general than as applied to a full amplitude: it can also be applied to individual terms in the string theory integrand. To understand the second issue, one needs to know the intimate relationship between string theory integrals and CHY integrals, as explained in ref.~\cite{Bjerrum-Bohr:2014qwa} (see also section 3 of ref.~\cite{Baadsgaard:2015voa}). The latter connection allows us to import monodromy relations of string theory in the $\alpha'\!\to\!0$ limit into CHY integrands. In this way we establish a broad class of general relations satisfied by CHY integrals, corresponding to real and imaginary parts of string theory monodromy relations. Taking the real part, we obtain identities that involve only the CHY integration variables. As might have been guessed, such identities are in fact simple algebraic identities of the kind obtained by, $e.g.$, partial fractioning. However, the identities corresponding to taking the imaginary part are highly non-trivial, mixing integration variables with generalized Mandelstam variables. In this way, integration variables can, figuratively speaking, be traded for momenta. In particular, such identities can be used to lower the order of the poles, thus rendering those integrals doable by means of the integration rules derived in refs.~\cite{Baadsgaard:2015voa,Baadsgaard:2015ifa,Baadsgaard:2015hia}. {This provides a step-by-step implementation of integration rules that can be used for any $n$-point amplitude, i.e. we start from the most complicated integrands and reduce them step-wise to simpler integrands until we only have integrands that can be evaluated.}

In this paper, we describe this application of string theory monodromy relations and how it can be applied as a powerful and systematic tool for analytically integrating the terms that appear in the CHY representation of Yang-Mills amplitudes. Surely these tools have much broader applications, but we consider Yang-Mills amplitudes as our primary example. In \mbox{section \ref{sec:review_of_chy_and_string}}, we review how Yang-Mills amplitudes are represented in CHY and string theory, and discuss the obstacles to direct analytic integration. The first obstacle is the fact that the CHY representation is not manifestly M\"{o}bius-invariant term-by-term; this is remedied in \mbox{section \ref{sec:review_of_chy_and_string}} where we describe a refinement of the CHY representation that is manifestly M\"{o}bius-invariant.  Even when every term is manifestly M\"{o}bius-invariant, however, the analytic rules for integration described in~\cite{Baadsgaard:2015voa,Baadsgaard:2015ifa,Baadsgaard:2015hia} can be obstructed by the appearance of integrands with what we will call `problematic $k$-tuples'. These include (and generalize) the higher-poles that can appear in individual terms in the CHY and string theory representations. In \mbox{section \ref{sec:monodromy_reductions}}, we describe how monodromy relations of string theory can be used to systematically eliminate these obstructions. We use these new rules to derive analytic formulae (via CHY) for Yang-Mills amplitudes involving as many as six gluons. These are given in detail in \mbox{Appendix \ref{appendix:explicit_formulae}}; these formulae have been verified against known results (e.g.\ using the package~\cite{Bourjaily:2010wh}), and are provided as a {\sc Mathematica} notebook included in this work's submission files on the {\tt arXiv}.

\section{Review and Refinement of CHY and String Amplitudes}\label{sec:review_of_chy_and_string}
In this section, we rapidly review the CHY and string theory representations of amplitudes in Yang-Mills theory, and briefly discuss the obstacles to analytic integration of the formulae that result. But prior to doing so, we must first refine the CHY representation in order make it manifestly M\"{o}bius-invariant term-by-term.

In the scattering equation formalism, the $n$-point gluon amplitude in Yang-Mills can be represented as follows~\cite{Cachazo:2013gna,Cachazo:2013hca,Cachazo:2013iea},
\eq{\mathcal{A}_n\equiv(-1)^{\lfloor n/2\rfloor}\!\int\!\!\Omega_{\text{CHY}}\,\,\frac{\text{Pf}\,'\Psi(z_i)}{\z12\z23\cdots\z{n}{1}},\label{chy_rep_of_ym_amps}}
where the integration measure $\Omega_{\text{CHY}}$ (which includes the scattering equation constraints) is given by:
\eq{\!\!\!\!\!\!\!\!\!\!\!\!\!\!\!\Omega_{\text{CHY}}\equiv\frac{d^nz}{\mathrm{vol}(SL(2)\!)}\prod_i\,\!'\delta(S_i)\equiv\!\z{r}{s}^2\z{s}{t}^2\z{t}{r}^2\prod_{\fwbox{25pt}{i\!\in\!\mathbb{Z}_n\!\backslash\{r,s,t\}}} dz_i\,\delta(S_i)\,,\label{definition_of_chy_measure}}
where the $\delta$-functions impose the scattering equations,
\eq{S_i\equiv\sum_{j\neq i}\frac{s_{ij}}{\z{i}{j}}=0,}
localizing the integration to simply a sum over the $(n-3)!$ solutions to $\{S_i\!=\!0\}$; also appearing in the integration measure (\ref{chy_rep_of_ym_amps}) is the {\it reduced} Pfaffian~\footnote{\footnotesize{Interestingly, we can here report on one further refinement; one has {\it always} the freedom to pick a different Pfaffian reduction for each occurring product of contracted polarization vectors in the amplitude. Although not employed here, this observation can be used to favour certain CHY integrations when deriving amplitude results.}} of the matrix $\Psi$ (that is, the Pfaffian of $\Psi^{ij}_{ij}$, obtained by deleting rows and columns $i,j$ from $\Psi$),
\eq{\text{Pf}\,'\Psi\equiv\frac{(-1)^{i+j}}{\z{i}{j}}\text{Pf}\big(\Psi^{ij}_{ij}\big),\quad \text{where}\quad\Psi\equiv\left(\begin{array}{c@{$\,\,\,\,\,\,\,\,$}c@{$\,\,\,$}}A&\fwboxR{0pt}{-}C^{\fwboxL{0pt}{T}}\\C&B\end{array}\right),}
where the components of $\Psi$ are given by the matrices,
\begin{equation}\begin{split}&\fwboxL{100pt}{A_{i\neq j}\equiv\frac{s_{ij}}{\z{i}{j}},}\fwboxL{100pt}{B_{i\neq j}\equiv\frac{\e{i}{j}}{\z{i}{j}},}\fwboxL{100pt}{C_{i\neq j}\equiv\frac{\ek{i}{j}}{\z{i}{j}},}\\
&\fwboxL{100pt}{A_{i=j}\equiv0,}\fwboxL{100pt}{B_{i=j}\equiv0,}\fwboxL{100pt}{C_{i=j}\equiv-\sum_{l\neq i}\frac{\ek{i}{l}}{\z{i}{l}}.}\end{split}\end{equation}
for which $s_{ij}\!\equiv\! 2k_i\!\cdot\!k_j$ and $\e{i}{j}\!\equiv\!2\,\epsilon_i\!\cdot\!\epsilon_j$ and $\ek{i}{j}\!\equiv\!\ekexpl{i}{j}$.

While correct, this representation does not provide a manifestly M\"{o}bius-invariant integrand for the amplitude because of the diagonal terms of the matrix $C$: these terms are not of uniform (nor correct) weight under M\"{o}bius transformations. This problem can be solved as follows. Let us make use of the (partial-fraction) identity,
\eq{-\frac{\ek{i}{l}}{\z{i}{l}}=\frac{\ek{i}{l}}{\z{a}{i}}+\frac{\ek{i}{l}\z{l}{a}}{\z{a}{i}\z{i}{l}}\quad\text{for}\quad i\neq a,}
to re-write the diagonal terms of the $C$-matrix,
\eq{C_{ii}=\sum_{l\neq i}\left(\frac{\ek{i}{l}}{\z{a}{i}}+\frac{\ek{i}{l}\z{l}{a}}{\z{a}{i}\z{i}{l}}\right)\Rightarrow\sum_{l\neq i,a}\frac{\ek{i}{l}\z{l}{a}}{\z{a}{i}\z{i}{l}}.}
Here, the RHS follows from gauge-invariance (and momentum conservation)---as the sum of the first terms is always proportional to $\ek{i}{i}$. Because the terms on the RHS have uniform weight of $z_i^{-2}$ under modular transformations, the reduced Pfaffian is guaranteed to be term-wise M\"{o}bius invariant. Thus, and for the sake of concreteness, we can replace the diagonal elements of the $C$-matrix by, for example,
\vspace{-2.5pt}\eq{C_{ii}\Rightarrow\left\{\begin{array}{l@{$\qquad$}r}\displaystyle\,\,\,\sum_{l=3}^n\,\,\frac{\ek{1}{l}\z{l}{2}}{\z{2}{1}\z{1}{l}}\,,&i=1\,,\\\displaystyle\,\,\,\sum_{\fwbox{0pt}{l\!\not\in\!\{1,i\}}}\,\,\frac{\ek{i}{l}\z{l}{1}}{\z{1}{i}\z{i}{l}}\,,&i>1\,.\end{array}\right.\vspace{-4.5pt}\label{refined_version_of_c_matrix}}
Throughout the rest of this work, whenever we speak of `the' terms in the CHY representation of the amplitude, we have made use of this form of the diagonal entries of the $C$-matrix---rendering the CHY representation term-wise, manifestly M\"{o}bius-invariant.

Another way to compute pure Yang-Mills field theory amplitudes is provided by superstring theory---see, $e.g.$, \mbox{ref.~\cite{GSW}}. Here the $n$-point field theory amplitude can be computed as the leading $\alpha'$ contribution to a set of ordered integrations {along the real axis}:\\[-14pt]
\begin{align}\mathcal{A}_{n}\!=\!&\lim_{\alpha'\!\to0}{\alpha'}^{(n-4)/2}\,\int\prod_{i=3}^{n-1}dz_i\,{\z12\z2n\z{n}{1}\over \prod_{i=1}^n \z{i}{i+1}}\!\!\int\!\!d^n\theta\,d^n\phi\prod_{i<j}(z_i-z_j-\theta_i\theta_j)^{\alpha'\s{i}{j}}\nonumber\\
&\hspace{25pt}\times\prod_{i<j}\exp\left[\frac{\sqrt{2\alpha'}(\theta_i\mi\theta_j)\big(\phi_i\ek{i}{j}\pl\phi_j\ek{i}{j}\big)}{\z{i}{j}} \right.\mi\left. \frac{\phi_i\phi_j\e{i}{j}}{\z{i}{j}}\mi\frac{\theta_i\theta_j\phi_i\phi_j\e{i}{j}}{\z{i}{j}^2}\right]\!.\hspace{-30pt}\\[-16pt]\nonumber\end{align}
The auxiliary Grassmann integrations over $\phi_i$ and $\theta_i$ automatically impose the multi-linearity condition on the amplitude in terms of the external polarization vectors $\epsilon_j^{\mu}$, just like the Pfaffian does in the CHY prescription. Explicit examples of using string theory to compute Yang-Mills amplitudes, including all the stringy corrections proportional to powers of $\alpha'$ can be found in~\cite{GSW} and in the impressive work by Medina, Brandt and Machado~\cite{Medina:2002nk} (at 5-point, see also~\cite{Machado:2004ya}), and by Oprisa and Stieberger~\cite{Oprisa:2005wu} (at 6-point). {The pure spinor formalism provides another method to derive such amplitudes using the Berends-Giele recursion procedure~\cite{Berends:1987me,Mafra:2011nv,Mafra:2015vca}\footnote{\footnotesize{We thank C. Mafra and O. Schlotterer for informing us, after the preprint of this paper was made public, of the link {\tt http://www.damtp.cam.ac.uk/user/crm66/SYM/pss.html}, where many explicit examples of amplitudes are provided.}}.} Once the Grassmann integrations have been performed, we are left with bosonic integrands with poles in the $z_i$ variables. Using integration by parts identities a bosonic integrand written solely in terms of single poles can be recovered~\cite{Bjerrum-Bohr:2014qwa}. Inserting the CHY $\delta$-function constraints into such a superstring integrand and taking the $\alpha'\!\to\!0$ limit one precisely recovers the CHY prescription~\cite{Cachazo:2013gna,Cachazo:2013hca,Cachazo:2013iea} for Yang-Mills theory. An alternative, string-like derivation of the CHY formalism uses the ambitwistor string~\cite{Mason:2013sva,Berkovits:2013xba,Adamo:2013tsa,Geyer:2015bja,Geyer:2015jch}.

In ref.~\cite{Baadsgaard:2015voa}, this match between ordered string theory integrations and the CHY prescription was exploited in several ways. It is instructive to see why certain string theory integration rules do not immediately carry over to CHY-type integrals, while others do. Let us start with string theory and the following generic $\phi^3$-type integral over ordered variables,
\vspace{-5pt}\eq{\hspace{-20pt}{\cal I}_n \!= \lim_{\alpha'\!\to 0}{\alpha'\hspace{1pt}}^{n-3}\int\prod_{i=3}^{n-1}dz_i\,\z12\z2n\z{n}{1}\prod_{\!\!1\leq i< j\leq n\!\!} |z_i-z_j|^{\alpha' s_{ij}}H(z)\,,\vspace{-8.5pt}}
where $H(z)$ consists of products of factors $\z{i}{j}^{-\ell}$ such that the whole integrand is $SL(2)$-invariant. Depending on the form of $H(z)$, the integral above, with the prefactor $(\alpha')^{n-3}$, may or may not be well defined. If the degree of divergence of the integral itself is stronger than $(\alpha')^{3-n}$ as $\alpha' \to 0$ the evaluation of ${\cal I}_n$ will require analytic continuation. In ref.~\cite{Baadsgaard:2015voa} such integrals were not considered. This is sufficient to provide, for example, all integration rules for scalar $\phi^3$-theory. Tellingly, it is precisely these ``simpler'' string theory integrals for which compact integration rules can be formulated and for which there is one-to-one translation table to CHY integrals, where the corresponding integrals instead are evaluated by means of the global residue theorem. When we turn to Yang-Mills theory in the CHY formalism a more general set of integrals appear, and we need integration rules for them. This is where monodromy provides a solution. By deforming contours in string theory the analytic continuation can be performed in a systematic manner, relating the result to string theory integrations that do not require analytic continuation. The latter can immediately be transcribed into alternative CHY representations of the original integrals, now with the bonus that the standard integration rules apply.

Although the integration rules derived in ref.~\cite{Baadsgaard:2015voa} are very powerful and exhaust all integrals that arise for $\phi^3$-theory, certain integrations that arise in the CHY formulation of Yang-Mills theory are not covered by these rules. In string theory, those integrals are not well-defined for $\alpha'$ near the origin, requiring analytical continuation. This makes it more complicated to deduce proper integration rules, and interestingly this is true also in the CHY formalism. Steps have recently been taken towards the formulation of such generalized CHY integration rules in refs.~\cite{Gomez:2016bmv,Huang:2016zzb}. In the next section we will present a systematic solution to this problem. But before doing so, let us first review the obstructions that arise for more general integrands---and how we can represent these diagrammatically.

\subsection{Graphical Representations of Integrands and Obstacles to Integration}\label{subsec:graphical_reps_and_integration_rules}\vspace{-6pt}

We can represent any CHY/string-theory integrand $H(z)$ constructed as products of factors of the form $\z{i}{j}$ graphically as a multi-graph with solid lines indicating factors that appear in the denominator (with multiplicity), and with dashed lines indicating factors in the numerator (with multiplicity). For example,
\eq{\hspace{-10pt}\fig{-26.05pt}{0.35}{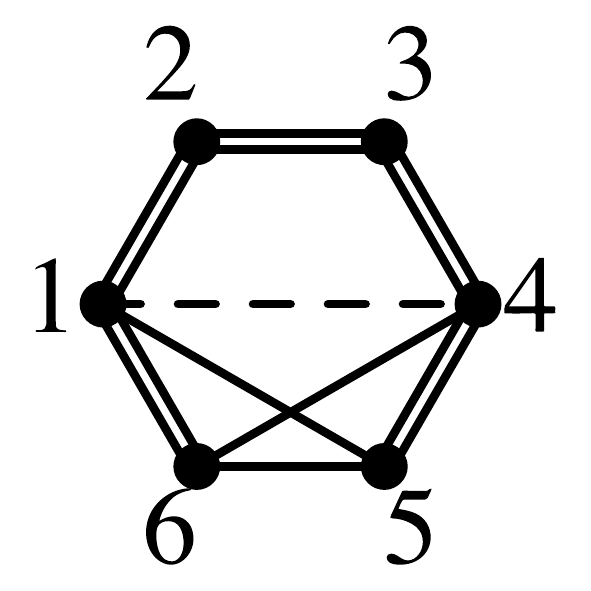}\!\Leftrightarrow\frac{\z14}{\z12^2\z23^2\z34^2\z45^2\z56\z16^2\z15\z46}.\nonumber}
To be completely clear throughout this work, we will always use the convention that every link $(ij)\!\Leftrightarrow\!\z{i}{j}$ that appears in the graph is taken to be {\it ordered}, with $i\!<\!j$. Thus, when we find it useful later on to discuss  `Parke-Taylor'-like factors $1/(\z12\cdots\z{n}{1})$, the reader should bear in mind that this would be represented graphically with a minus sign: e.g.,
\eq{\hspace{-10pt}\fig{-26.05pt}{0.35}{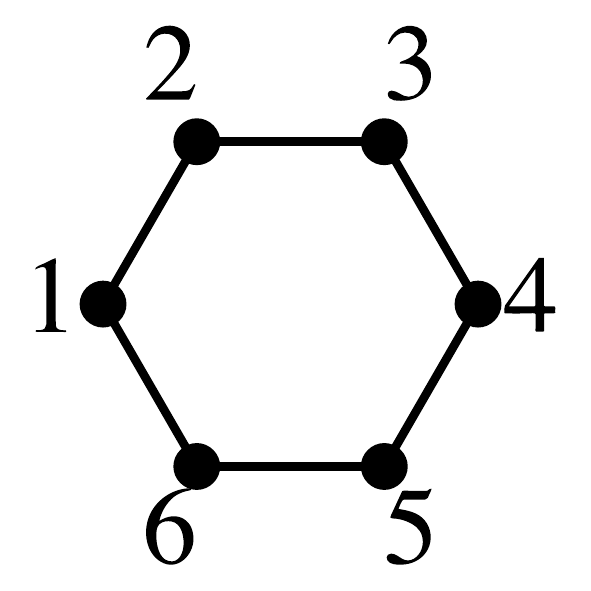}\!\Leftrightarrow\frac{1}{\z12\z23\z34\z45\z56\z16}={\color{hred}-}PT(1,\!2,\!3,\!4,\!5,\!6).\nonumber}

We need not review the combinatorial rules for analytic integration described in ref.~\cite{Baadsgaard:2015voa}. But for our purposes it will be important to emphasize that these rules necessitate that for every $k$-element subset of particle labels $\tau$, there exists no more than $2k\mi2$ factors $\z{i}{j}$ in the denominator between elements $\{i,j\}\!\subset\!\tau$ (counting factors in the numerator negatively). Subsets $\tau$ that {\it do not} meet this criterion will be called `problematic $k$-tuples'. When an integrand is free of problematic $k$-tuples, then the integration rules described in ref.~\cite{Baadsgaard:2015voa} apply, providing an analytic expression for the result of integration against the CHY measure.

Both the six-point integrands drawn above are free of problematic $k$-tuples, and hence can be integrated analytically without difficulty. Perhaps the simplest example of a graph with a problematic $k$-tuple appears for $4$ particles:
\eq{\hspace{-10pt}\fig{-18.65pt}{0.3}{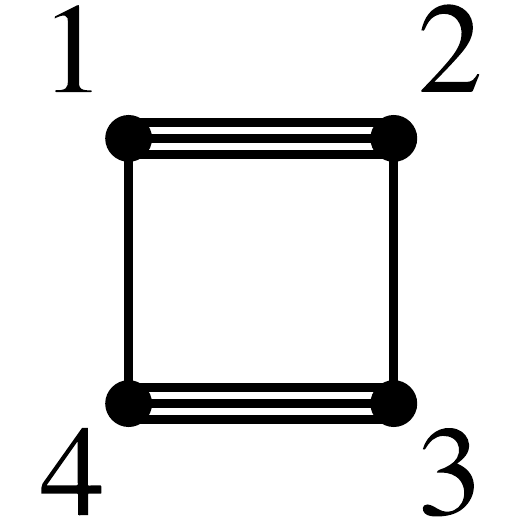}\!\Leftrightarrow\frac{1}{\z12^3\z23\z34^3\z14},\label{4pt_example_triple_line_graph}}
for which the $2$-tuple $\tau\!\equiv\!\{1,2\}$ is problematic because there are more than 2 factors of $\z{1}{2}$ in the denominator. We could also describe the subset $\{3,4\}$ as problematic, but subsets should be considered equivalent to their complements {so it is sufficient to consider only $\tau\!\equiv\!\{1,2\}$.} The existence of a problematic $2$-tuple is always indicated by a triple-line in the diagrammatic representation of the integrand.

A more intricate example of an integrand with problematic $k$-tuples would be the following:
\eq{\hspace{-10pt}\fig{-26.05pt}{0.35}{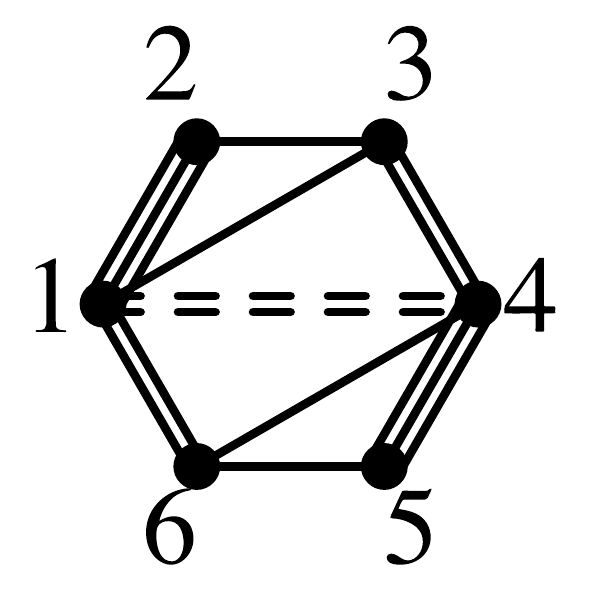}\!\Leftrightarrow\frac{\z14^2}{\z12^3\z23\z34^2\z45^3\z56\z16^2\z13\z46}.\nonumber}
This integrand has four problematic $k$-tuples: $\{1,2\}, \{4,5\}, \{1,2,3\},$ and $\{1,2,6\}$.

In the next section we will describe how integrands such as these with problematic $k$-tuples can systematically be expanded using monodromy relations into a sum of integrands without problematic $k$-tuples, allowing us to use the combinatorial rules of ref.~\cite{Baadsgaard:2015voa} to express the result of their integration analytically.

\section{Integrand-Level Monodromy Relations and Reduction}\label{sec:monodromy_reductions}\vspace{-6pt}

As reviewed above, the two primary obstacles to obtaining analytic formulae for scattering amplitudes using the scattering equation formalism are the non-manifest M\"{o}bius-invariance of individual terms---solved in our refined formulation---and the appearance of integrands such as (\ref{4pt_example_triple_line_graph}) that have problematic $k$-tuples. To illustrate this, let us consider the terms that appear in the (refined) CHY representation of the $4$-particle tree-amplitude. Using (\ref{chy_rep_of_ym_amps}) with $C$ defined according to (\ref{refined_version_of_c_matrix}), picking $\{i,j\}\!=\!\{1,2\}$ for the projection to the reduced Pfaffian, and extracting the coefficients of cyclic classes (mod duplication), the amplitude is expressed as follows,\\[-12pt]
\eq{\mathcal{A}_4=\alpha_1\,\e{1}{2}\e{3}{4}+\alpha_2\,\e{1}{3}\e{2}{4}+\beta_1\,\e{1}{2}+\beta_2\,\e{1}{3}+\text{{\it distinct} cyclic},\label{four_point_tree_seeds_in_chy}}
where the coefficients are given by:
\begin{equation}\begin{split}\alpha_1&\equiv s_{12}\!\!\fig{-18.65pt}{0.3}{4pt_term_1}, \fwboxR{162.15pt}{\quad\qquad\qquad \alpha_2\equiv - s_{12}\!\!\fig{-18.65pt}{0.3}{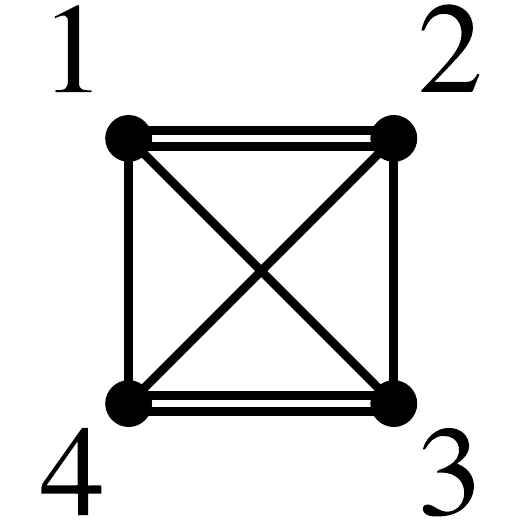},}\\
\beta_1&\equiv\phantom{-} \ek32 \ek41\!\!\fig{-18.65pt}{0.30}{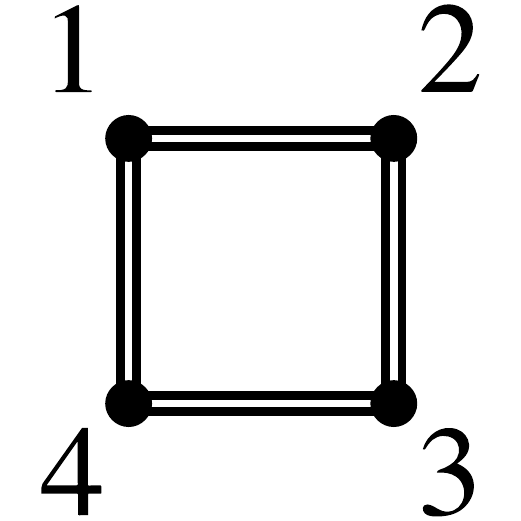}-\ek31 \ek42\!\!\fig{-22.25pt}{0.30}{4pt_term_2},\\
\beta_2&\equiv -\ek23 \ek41\!\!\fig{-18.65pt}{0.30}{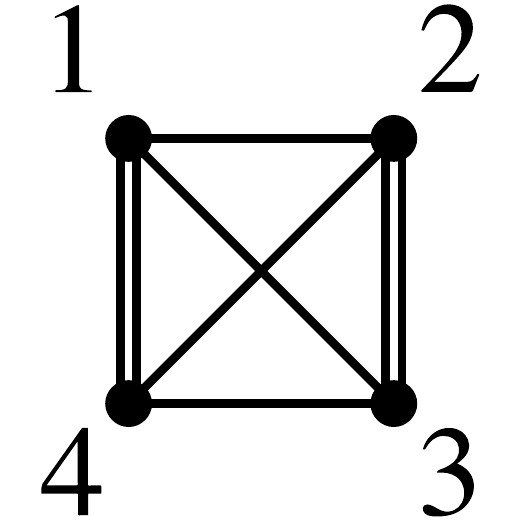}-\ek21 \ek43\!\!\fig{-18.65pt}{0.30}{4pt_term_2}.\end{split}\label{four_point_cyclic_seed_graphs}\end{equation}
Of these, all but $\alpha_1$ can be integrated immediately via the rules of ref.~\cite{Baadsgaard:2015voa}:
\eq{\hspace{-20pt}\fig{-18.65pt}{0.30}{4pt_term_2}\!\!=-\frac{1}{s_{12}}, \qquad \fig{-18.65pt}{0.30}{4pt_term_3}\!\!=-\frac{1}{s_{23}},\qquad \fig{-18.65pt}{0.30}{4pt_term_4}\!\!=-\left(\frac{1}{s_{12}}+\frac{1}{s_{23}}\right),}
from which we see that $\alpha_2\!=\!1$,
\eq{\hspace{-30pt}\beta_1=\frac{\ek31\ek42s_{23}\!+\!\ek32\ek41s_{13}}{s_{12}s_{23}},\quad\beta_2=\frac{\ek21\ek43s_{23}\!+\!\ek23\ek41s_{12}}{s_{12}s_{23}}.}

While the CHY integrand appearing in the coefficient $\alpha_1$ is M\"{o}bius invariant, it cannot be integrated analytically according to the rules of ref.~\cite{Baadsgaard:2015voa} because of the cubic powers $\z12^3$ and $\z34^3$ appearing in the denominator (represented as triple lines in the figure). As described above, these indicate the existence of the problematic $2$-tuple $\{1,2\}$.

Let us now describe how monodromy relations of string theory can remedy this situation---lowering the degree of poles in the diagram (\ref{4pt_example_triple_line_graph}). The basic idea is a simple one. Viewing the integrand (\ref{4pt_example_triple_line_graph}) in string theory, monodromy tells us how to exchange one integration region with another while carefully deforming the contour around branch points. Effectively, this results in complex phases (determined by the Koba-Nielsen factor) attached to the integrand:
\vspace{-5pt}\begin{align}\hspace{-10pt}0=&\int\limits_{-\infty}^0\!\!dz {H}( z) (-z)^{\alpha' \s{1}{2}}(1-z)^{\alpha' \s{2}{3}}\label{basic_monodromy_relation_sketch}\\
&\hspace{-0pt}+e^{i\alpha' \s{1}{2}} \int\limits_{0}^1\!\!dz \, {H}( z) (z)^{\alpha' \s{1}{2}}(1-z)^{\alpha' \s{2}{3}}+e^{i\alpha' (\s{1}{2}+\s{2}{3})}\int\limits_{1}^\infty\!\!dz \, {H}( z) (z)^{\alpha' \s{1}{2}}(z-1)^{\alpha' \s{2}{3}}\,.\nonumber\end{align}
{Let us introduce a convenient graphical notation. A line between two points  $i<j$ represents a factor ${1\over \z{i}{j}}$ both with
respect to the string theory and the CHY measures. Applied to the case of eq.~\ref{4pt_example_triple_line_graph}, the above relation then becomes a three-term identity:}
\eq{\fwboxL{290pt}{\fwboxL{70pt}{0=\!\!\!\fig{-18.65pt}{0.3}{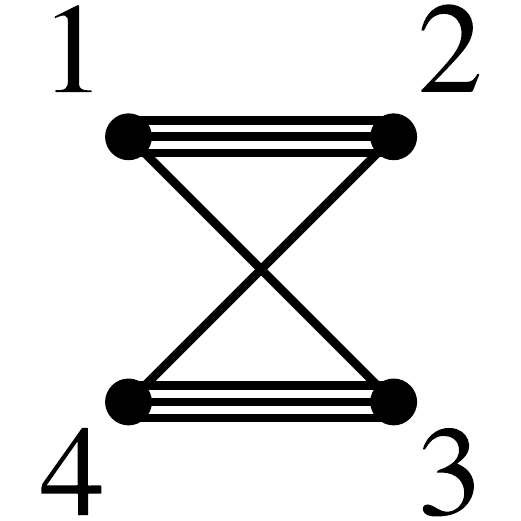}}\!\!\!\fwboxR{100pt}{+\;\,\,\;e^{i\alpha's_{12}}\!\!\!\fig{-18.65pt}{0.3}{4pt_term_1}}\!\!\!\fwboxR{140pt}{-\;\;\;\;e^{i\alpha'(s_{12}+s_{23})}\!\!\!\fig{-18.65pt}{0.3}{4pt_term_2}}}~\,.\label{four_point_example_monodromy_relation_full}}
Here, the minus sign appearing in the relation above is really due to our {\it convention} for how to order the denominators of the factors corresponding to the diagrams. Such a relation naturally splits up into real and imaginary parts~\cite{Mafra:2011kj,BjerrumBohr:2009rd,BjerrumBohr:2010zs,Cachazo:2012uq}, yielding:
\eq{\begin{split}&\fwboxL{290pt}{\fwboxL{70pt}{0=\!\!\!\fig{-18.65pt}{0.3}{4pt_term_5}}\!\!\!\fwboxR{100pt}{+\,\cos\!\big(\alpha's_{12}\big)\!\!\!\!\!\!\fig{-18.65pt}{0.3}{4pt_term_1}}\!\!\!\fwboxR{140pt}{-\,\,\cos\!\big(\alpha'(s_{12}\!+\!s_{23})\big)\!\!\!\!\!\!\fig{-18.65pt}{0.3}{4pt_term_2}}}~\,,\\
&\fwboxL{290pt}{\fwboxL{70pt}{}\!\!\!\fwboxR{100pt}{0=\,\sin\!\big(\alpha's_{12}\big)\!\!\!\!\!\!\fig{-18.65pt}{0.3}{4pt_term_1}}\!\!\!\fwboxR{140pt}{-\,\,\sin\!\big(\alpha'(s_{12}\!+\!s_{23})\big)\!\!\!\!\!\!\fig{-18.65pt}{0.3}{4pt_term_2}}}~\,.\end{split}\label{four_point_example_monodromy_relation_parts}}
These identities are the analogs of KK~\cite{KK} and BCJ~\cite{BCJ} relations, respectively. Note that the first relation (the real part) involves two diagrams both with triple lines. The identity holds, of course; but it is not the one that will prove useful to us here. The relation following from the imaginary part, however, is far more interesting: it relates a diagram with a triple line (a problematic $2$-tuple) to one without. As we are only interested in the leading contribution as $\alpha'\!\to\!0$, this identity becomes,
\begin{equation}\begin{split}\fig{-18.65pt}{0.3}{4pt_term_1}&=\frac{s_{12}\!+\!s_{23}}{s_{12}}\!\!\!\fig{-18.65pt}{0.3}{4pt_term_2}=-\frac{s_{12}\!+\!s_{23}}{s_{12}^2}=\frac{s_{13}}{s_{12}^2}.\end{split}\end{equation}

Using this, we see that $\alpha_1$ given in (\ref{four_point_cyclic_seed_graphs}) is simply equal to $s_{13}/s_{12}$. Thus, we have found analytic expressions for all the terms needed to express the amplitude. Putting everything together, we have:
\begin{equation}\begin{split}\hspace{-5pt}\mathcal{A}_4=&\fwboxR{30pt}{\Big[\e13\e24}+\frac{1}{s_{12}}\Big(\e12\e34s_{13}\pl\e12\big(\ek31\ek42\pl\ek32\ek41\big)\pl\e13\ek21\ek43\Big)\\
&\fwbox{30pt}{}+\frac{1}{s_{23}}\Big(\e12\ek32\ek41\pl\e13\ek23\ek41\Big)\Big]+\text{{\it distinct} cyclic}.\end{split}\label{full_four_point_amplitude_cyclic_seeds}\end{equation}

Going to higher multiplicity, the terms generated in the CHY representation increasingly involve problematic $k$-tuples. For $n\!=\!5$, for example, a direct expansion of the CHY representation (\ref{chy_rep_of_ym_amps}) (using the refined $C$-matrix and projecting to the reduced Pfaffian with $\{i,j\}\!=\!\{1,2\}$---for the sake of concreteness) generates an expansion involving 26 distinct CHY integrals to evaluate. Of these, 17 are free of problematic $k$-tuples and therefore can be integrated directly using the tools of ref.~\cite{Baadsgaard:2015voa}. The diagrams that have problematic $k$-tuples include, for example,
\eq{\fig{-26.05pt}{0.35}{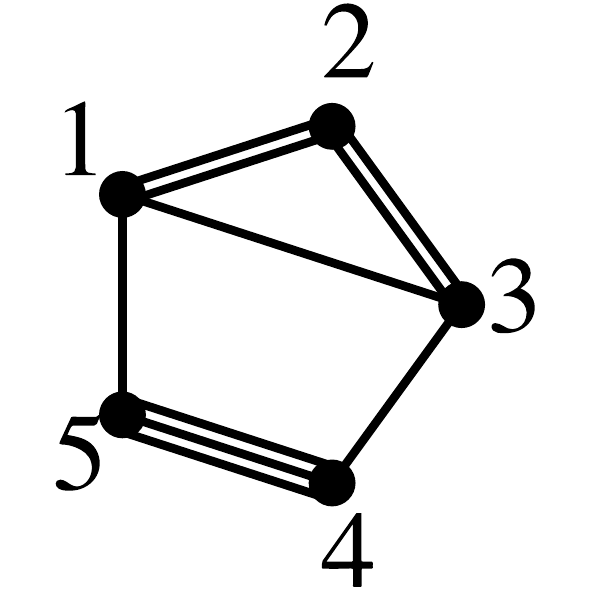}\qquad\fig{-26.05pt}{0.35}{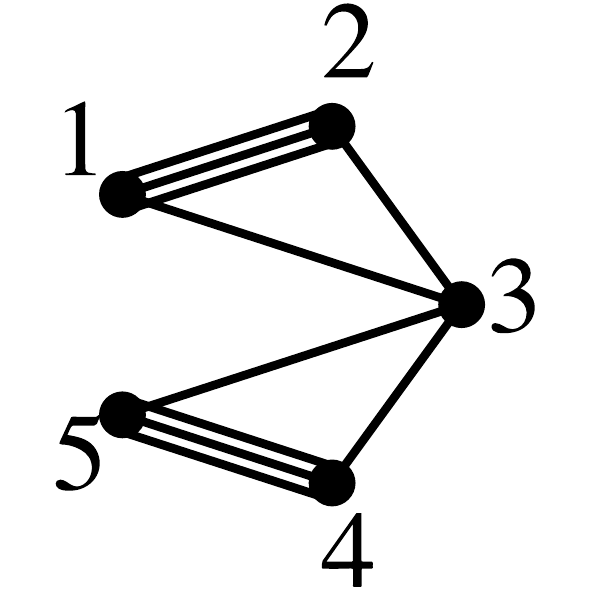}\qquad\fig{-26.05pt}{0.35}{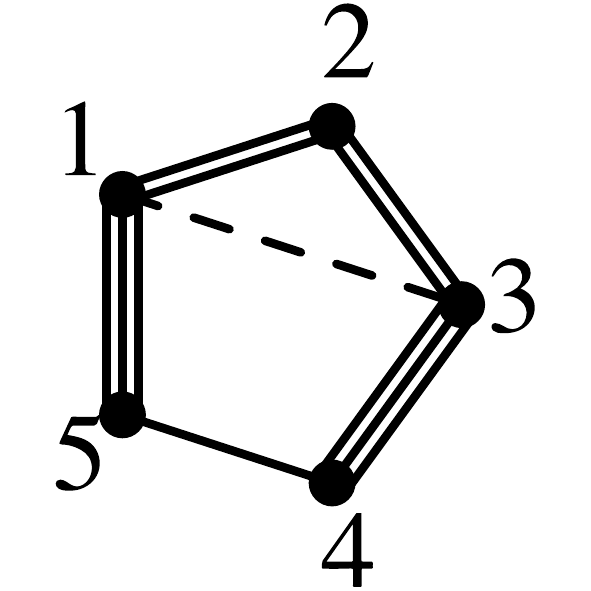}\label{five_point_exempli_with_problematic_tuples}}

Like for $n\!=\!4$, the only problematic $k$-tuples are $2$-tuples when $n\!=\!5$ (simply because subsets are considered equivalent to their complements). Thus, we should be able to use the same strategy as above to compute such terms analytically.

\subsection{Systematic Elimination of Problematic $2$-Tuples}\label{sec:two_tuple_reductions}\vspace{-6pt}

Let us now describe how problematic $2$-tuples can be systematically eliminated through a natural generalization of the identity (\ref{basic_monodromy_relation_sketch}). This will allow us to analytically integrate all the terms appearing the $5$-particle amplitude.

In order to describe the generalization of (\ref{basic_monodromy_relation_sketch}) to higher multiplicity, it will be useful to define the notation
\eq{PT(1,2,\ldots,n)\equiv\frac{1}{\z12\z23\z34\cdots\z{n}{1}},}
(motivated by analogy to the structure of the Parke-Taylor amplitude,~\cite{Parke:1986gb}). In the CHY representation of Yang-Mills amplitudes (\ref{chy_rep_of_ym_amps}), every term in the $n$-particle amplitude is manifestly proportional to $PT(1,\ldots,n)$. But introducing this notation here will allow us to deal with more general Hamiltonian cycles {(a path through a graph that passes through all vertices exactly once)} appearing in the integrands in which we are interested.

{It is straightforward to see that the generalized BCJ-type identity from the imaginary part of the basic monodromy relation (\ref{basic_monodromy_relation_sketch}) (at leading order in $\alpha'$) is the identity:}
\vspace{-2.5pt}\eq{0=s_{12}PT(1,{\color{hred}2},\ldots,n)+\sum_{k=3}^{n-1}(s_{12}\!+\!s_{2(3\cdots k)})PT(1,\ldots,k,{\color{hred}2},k\text{+}1,\ldots,n).\label{general_2_tuple_relation}\vspace{-2.5pt}}
{as anticipated from (\ref{four_point_example_monodromy_relation_full}) and (\ref{four_point_example_monodromy_relation_parts}).}
Here, we have introduced the notation $s_{a(b\cdots c)}\!\equiv\!s_{ab}+\ldots+s_{ac}$ for the sake of concision. Just to be clear, this is not an `identity' among CHY {\it integrands}, but an identity {\it after} integration against the scattering equation constraints. We will give an alternate, direct proof of this identity in \mbox{Appendix \ref{appendix:monodromy_elaboration}}. 
Dividing by the Parke-Taylor pre-factor in the leading term of (\ref{general_2_tuple_relation}), we can re-write this identity in terms of cross-ratios constructed from the $z_i$'s:
\eq{1=-\sum_{k=3}^{n-1}\left(\frac{s_{12}\!+\!s_{2(3\cdots k)}}{s_{12}}\right)\frac{\z12\z23\z{k}{k\text{+}1}}{\z13\z{k}{2}\z{2}{k\text{+}1}}\,.\label{general_2_tuple_relation_bos_form}}

Importantly, multiplication of any CHY integrand by (\ref{general_2_tuple_relation_bos_form}) will result in sum of integrands with a reduced power of $\z12$ appearing in the denominator. For example, an integrand with the problematic $2$-tuple $\{1,2\}$ (corresponding to a factor of $1/\z12^3$) will be expanded into a sum of terms proportional to $1/\z12^2$---free of the problematic $2$-tuple. Thus, the identity systematically eliminates the problematic $2$-tuple $\{1,2\}$. This motivates us to label this identity as follows:
\eq{\mathrm{Id}_{\{1,2\}}\equiv-\sum_{k=3}^{n-1}\Big(\frac{s_{12}+s_{2(3\cdots k)}}{s_{12}}\Big)\frac{PT(1,\ldots,k,2,k\text{+}1,\ldots,n)}{PT(1,2,\ldots,n)}=1.\label{general_2_tuple_id}}
(Strictly speaking, this identity also depends on an overall cyclic ordering---through the appearance of $PT(1,2,\ldots,n)$ in the denominator of (\ref{general_2_tuple_id}). However, any permutation $\sigma\!\in\!\mathfrak{S}_n$ of labels $(1,2,\ldots,n)\!\to\!(\sigma_1,\sigma_2,\ldots,\sigma_n)$ such that $\{1,2\}\!\subset\!\{\sigma_1,\sigma_2\}$ would achieve the elimination of the bad $2$-tuple $\{1,2\}$. Usually there is a natural choice for the cyclic ordering as every graph (including those generated by multiple iterations of identities such as (\ref{general_2_tuple_id})) will involve a Parke-Taylor prefactor; when this is the case, use of this identity will not generate any new factors in the numerator. In our examples below, the `natural' ordering will always be taken.)

This notation should be fairly intuitive: for any CHY integration with a problematic $2$-tuple $\tau$, multiplication by $\mathrm{Id}_{\tau}$ will result in a sum of terms without the problematic $2$-tuple. This can be done iteratively, leading to a systematic elimination of all problematic $2$-tuples, allowing us to obtain analytic  expressions for these terms using the integration rules of ref.~\cite{Baadsgaard:2015voa}.

As described above, for $n\!=\!5$ the only possible bad $k$-tuples are $2$-tuples. Thus, the procedure described above should suffice to systematically evaluate terms such as those in (\ref{five_point_exempli_with_problematic_tuples})---examples relevant to the $5$-particle amplitude. The first of the examples in (\ref{five_point_exempli_with_problematic_tuples}) contains only a single problematic $2$-tuple---namely, $\{4,5\}$. Thus, it can be evaluated by a single application of $\mathrm{Id}_{\{4,5\}}$:
\vspace{-5pt}\begin{equation}\begin{split}\hspace{-45.5pt}\mathrm{Id}_{\{4,5\}}\hspace{-5pt}\fig{-26.05pt}{0.35}{5pt_exempli_1_0}&=\!\frac{s_{45}\!+\!s_{15}}{s_{45}}\!\!\!\fig{-26.05pt}{0.35}{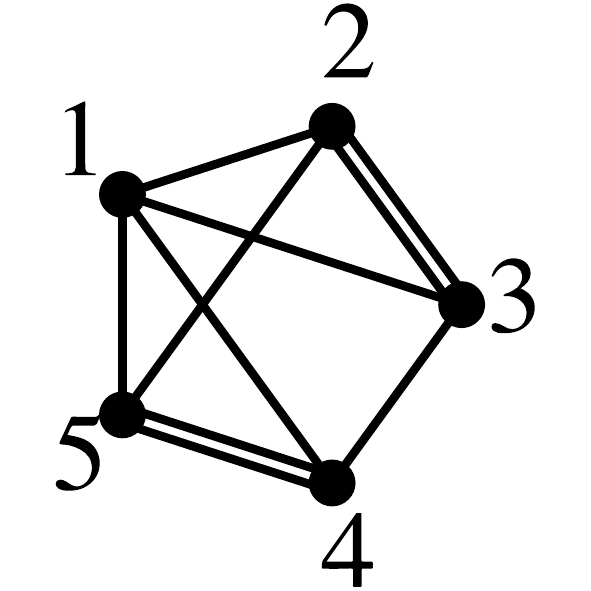}\!+\!\frac{s_{45}\!+\!s_{(12)5}}{s_{45}}\!\!\!\fig{-26.05pt}{0.35}{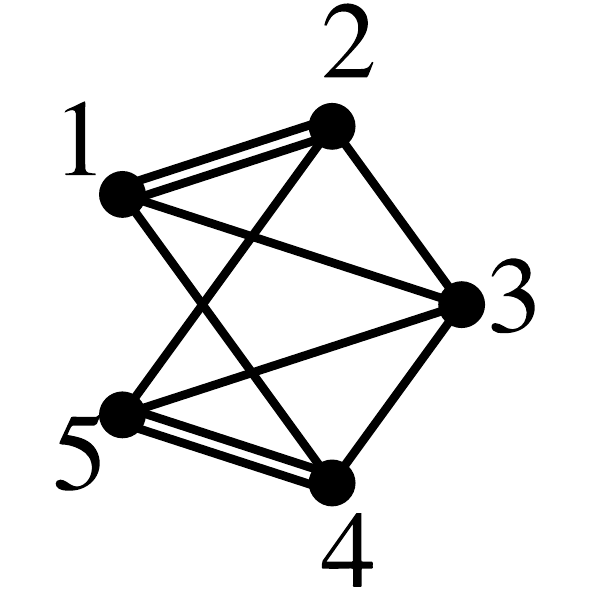}\\&=\frac{1}{s_{45}^2}\left(\frac{s_{45}\!+\!s_{15}}{s_{23}}-\frac{s_{35}}{s_{12}}\right).\label{five_particle_example_1}\end{split}\vspace{-5pt}\end{equation}
(We remind the reader that any unusual signs appearing above follow from the convention that all the links $(ij)\!\Leftrightarrow\!\z{i}{j}$ that appear in the graph are {\it ordered}: $i\!<\!j$.)

The other two examples are more involved, as each has two distinct problematic $2$-tuples. Nevertheless, repeated application of the identity (\ref{general_2_tuple_id}) will always result in an expansion into terms without problematic $2$-tuples. For the first, we find:
\begin{align}\hspace{-28.5pt}\hspace{-5pt}\mathrm{Id}_{\{4,5\}}\mathrm{Id}_{\{1,2\}}\hspace{-5pt}\fig{-26.05pt}{0.35}{5pt_exempli_2_0}\hspace{-5pt}&=-\frac{s_{(12)3}s_{(34)5}}{s_{12}s_{45}}\!\!\!\fig{-26.05pt}{0.35}{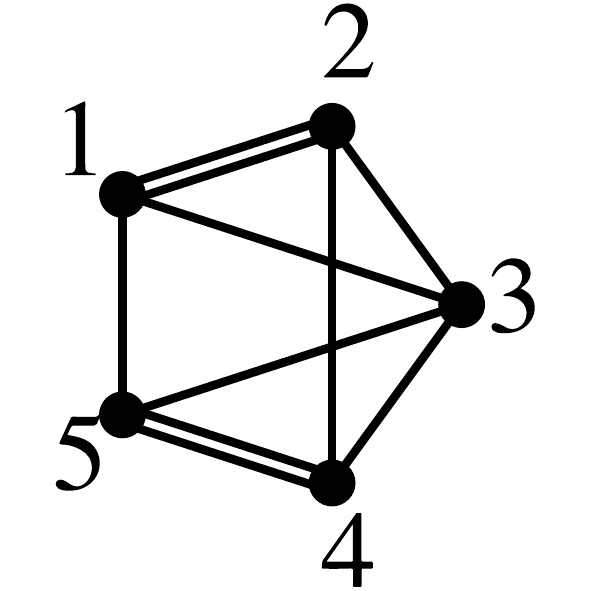}\!\!\!-\!\frac{s_{(12)3}s_{25}}{s_{12}s_{45}}\!\!\!\fig{-26.05pt}{0.35}{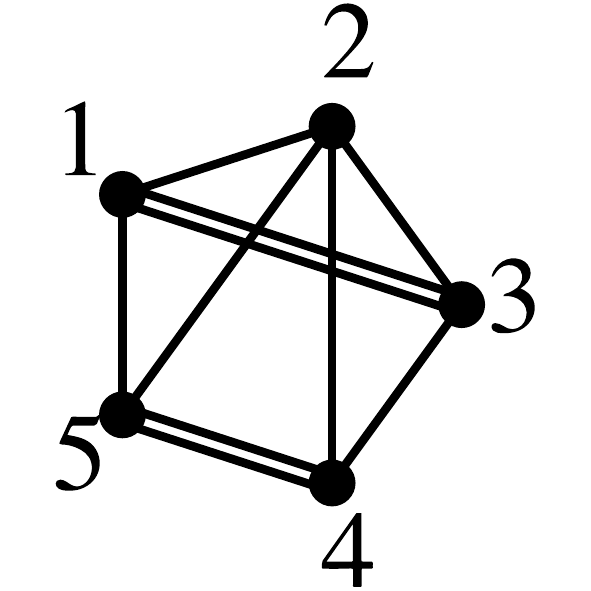}\!\!\!-\!\frac{s_{25}}{s_{12}}\!\!\!\fig{-26.05pt}{0.35}{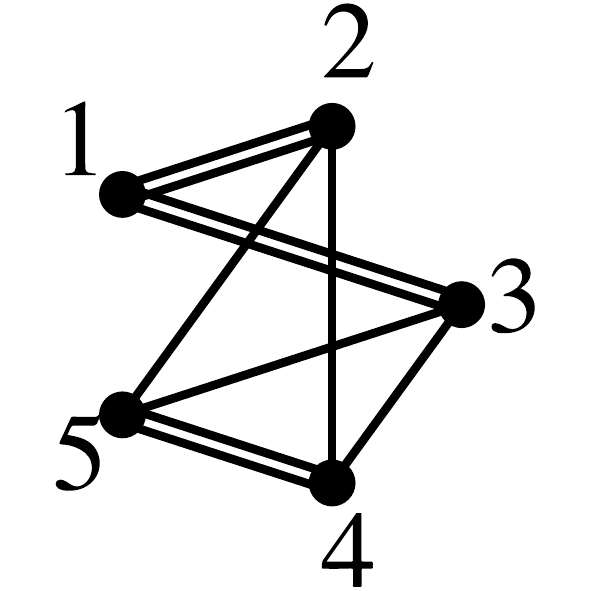}\!\!\hspace{-40pt}\nonumber\\&=\frac{s_{(12)3}}{s_{12}s_{13}s_{45}}\left(\frac{s_{13}s_{(34)5}}{s_{12}s_{45}}-\frac{s_{25}}{s_{45}}+\frac{s_{25}}{s_{12}}\right).\label{five_particle_example_2}\end{align}
And for the last example of (\ref{five_point_exempli_with_problematic_tuples}), we have:
\begin{align}\hspace{-38.5pt}\hspace{-5pt}\mathrm{Id}_{\{5,1\}}\mathrm{Id}_{\{3,4\}}\hspace{-5pt}\fig{-26.05pt}{0.35}{5pt_exempli_3_0}\hspace{-5pt}&=\frac{s_{1(25)}s_{4(35)}}{s_{15}s_{34}}\!\!\!\fig{-26.05pt}{0.35}{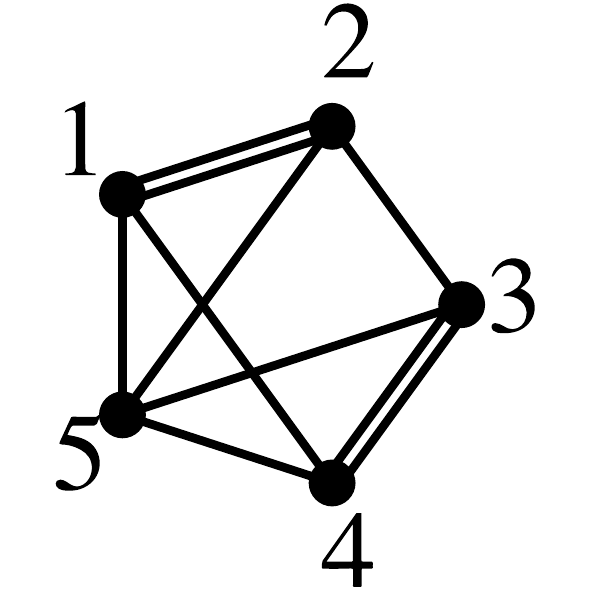}\!\!\!-\!\frac{s_{1(25)}s_{24}}{s_{15}s_{34}}\!\!\!\fig{-26.05pt}{0.35}{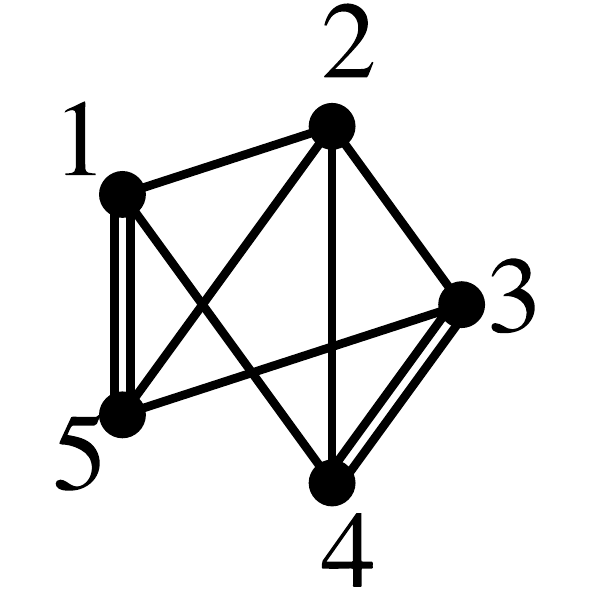}\!\!\!-\!\frac{s_{14}}{s_{15}}\!\!\!\fig{-26.05pt}{0.35}{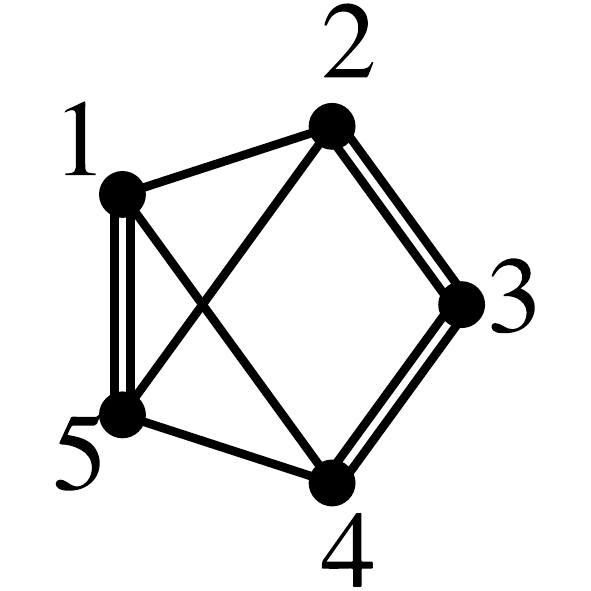}\!\!\hspace{-40pt}\nonumber\\&=\frac{1}{s_{15}s_{34}}\left(\frac{s_{1(25)}s_{4(35)}}{s_{12}s_{34}}-\frac{s_{1(25)}s_{24}}{s_{15}s_{34}}-\frac{s_{3(24)}s_{14}}{s_{15}s_{23}}\right).\label{five_particle_example_3}\end{align}

In these examples involving multiple iterations of identities, the expressions above should be understood somewhat suggestively: after applying $\mathrm{Id}_{\{1,2\}}$ to the example in (\ref{five_particle_example_2}), each term generated will have a different `preferred' Parke-Taylor ordering---and hence, different preferred orderings for the subsequent application of $\mathrm{Id}_{\{4,5\}}$. Moreover, not all the terms generated by application $\mathrm{Id}_{\{1,2\}}$ require further expansion: the rightmost term in the first line of (\ref{five_particle_example_2}) is already free of problematic $2$-tuples and hence can be directly integrated analytically.

We have made use of the general identity (\ref{general_2_tuple_id}) to evaluate every term generated in the CHY representation of the $5$-particle amplitude. The explicit result has been given in \mbox{Appendix \ref{appendix:5_pt_formula}}.

Beyond $n\!=\!5$, however, integrands can involve higher-order problematic $k$-tuples. In general, the terms in the $n$-point amplitude can have problematic tuples with $k\!\leq\!\lfloor n/2\rfloor$. Thus, the identities (\ref{general_2_tuple_id}) require generalization. Conveniently, the obvious generalization---to BCJ-like identities with higher-order shuffles---works. We now describe how this works in detail.

\subsection{General Monodromy Reductions: Eliminating Problematic $k$-Tuples}\label{sec:general_reductions}\vspace{-6pt}

The complete generalization of the monodromy relations (\ref{general_2_tuple_relation}) can be written in the following way:\footnote{A derivation of the relation can be found in~\cite{Chen:2011jxa}.}
\eq{0=\sum_{\fwbox{30pt}{\hspace{100pt}\sigma\!\in\!\big(\{2,\ldots,k\}\!\shuffle\!\{k\text{+}1,\ldots,n\,\text{-}1\}\big)}}PT(1,\sigma_1,\ldots,\sigma_{n\,\text{-}2},n)\Big(s_{1\cdots k}+\sum_{\fwbox{0pt}{\{i,j\}|\sigma_i\!>\!\sigma_j}}s_{\sigma_i\,\sigma_j}\Big).\label{general_k_tuple_relation}}
Here, $\{2,\ldots,k\}\shuffle\{k\text{+}1,\ldots,n\,\text{-}1\}$ denotes the set of all `shuffles' of the sets $\{2,\ldots,k\}$ and $\{k\text{+}1,\ldots,n\,\text{-}1\}$---that is, all permutations that preserve the relative ordering of the sets. It may be useful to give a concrete example. When $n\!=\!6$ and $k\!=\!3$, (\ref{general_k_tuple_relation}) becomes the BCJ-like identity:
\begin{equation}\begin{split}\hspace{-00pt}0=&\phantom{+\,}PT(1,\!{\color{hred}2},\!{\color{hred}3},\!{\color{hblue}4},\!{\color{hblue}5},\!6)s_{123}\!+\!PT(1,\!{\color{hred}2},\!{\color{hblue}4},\!{\color{hred}3},\!{\color{hblue}5},\!6)(s_{123}\!+\!s_{34})\\
&\!+\!PT(1,\!{\color{hred}2},\!{\color{hblue}4},\!{\color{hblue}5},\!{\color{hred}3},\!6)(s_{123}\!+\!s_{3(45)})\!+\!PT(1,\!{\color{hblue}4},\!{\color{hred}2},\!{\color{hred}3},\!{\color{hblue}5},\!6)(s_{123}\!+\!s_{(23)4})\\
&\!+\!PT(1,\!{\color{hblue}4},\!{\color{hred}2},\!{\color{hblue}5},\!{\color{hred}3},\!6)(s_{123}\!+\!s_{(23)4}\!+\!s_{35})\!+\!PT(1,\!{\color{hblue}4},\!{\color{hblue}5},\!{\color{hred}2},\!{\color{hred}3},\!6)(s_{123}\!+\!s_{(23)(45)}).
\end{split}\end{equation}

Because we are always interested in using these identities to eliminate one of the terms (that involving the identity element of the shuffle), it is natural to rewrite (\ref{general_k_tuple_relation}) slightly as follows:
\eq{\hspace{-20pt}0=s_{1\cdots k}PT(1,2,\ldots,n)+\!\!\!\sum_{\fwbox{30pt}{\hspace{100pt}\sigma\!\in\!\big(\{2,\ldots,k\}\widetilde{\shuffle}\{k\text{+}1,\ldots,n\,\text{-}1\}\big)}}PT(1,\sigma_1,\ldots,\sigma_{n\,\text{-}2},n)\Big(s_{1\cdots k}+\sum_{\fwbox{0pt}{\{i,j\}|\sigma_i\!>\!\sigma_j}}s_{\sigma_i\,\sigma_j}\Big),}
where here, $\widetilde{\shuffle}$ is defined to be the set of shuffles {\it excluding the identity}. This leads to the new set of monodromy relations, naturally generalizing those defined in (\ref{general_2_tuple_id}):
\eq{\hspace{-30pt}\mathrm{Id}_{\{1,\ldots,k\}}\!\equiv\!\frac{-1}{PT(1,\ldots,n)s_{1\cdots k}}\!\!\!\sum_{\fwbox{30pt}{\hspace{100pt}\sigma\!\in\!\big(\{2,\ldots,k\}\widetilde{\shuffle}\{k\text{+}1,\ldots,n\,\text{-}1\}\big)}}\!\!\!PT(1,\sigma_1,\ldots,\sigma_{n\,\text{-}2},n)\Big(s_{1\cdots k}+\sum_{\fwbox{0pt}{\{i,j\}|\sigma_i\!>\!\sigma_j}}s_{\sigma_i\,\sigma_j}\Big)\!=\!1.\label{general_k_tuple_id}}

As before, it is easy to see that application of $\mathrm{Id}_{\tau}$ will eliminate any problematic $k$-tuple $\tau$. To illustrate the use of these generalized monodromy relations, consider the evaluation of a contribution to the $6$-point amplitude with a single problematic $3$-tuple $\{1,2,3\}$: through multiplication by $\mathrm{Id}_{\{1,2,3\}}$ we find,
\begin{align}\hspace{-5.5pt}\fig{-26.05pt}{0.35}{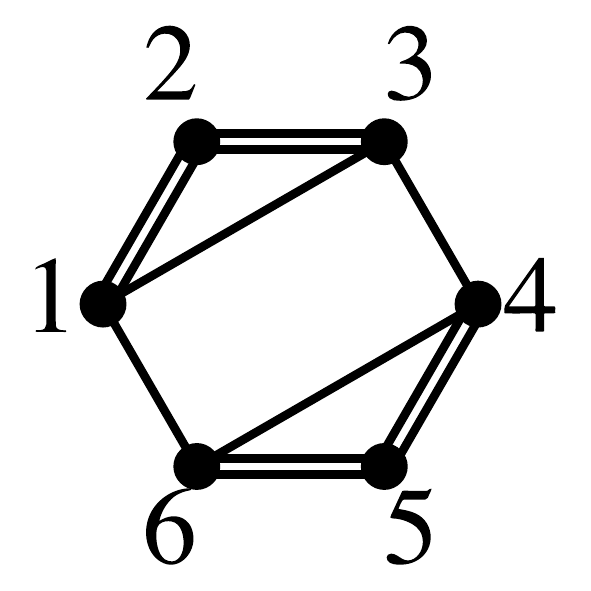}&=\phantom{+}\!\frac{s_{123}\!+\!s_{34}}{s_{123}}\!\!\!\fig{-26.05pt}{0.35}{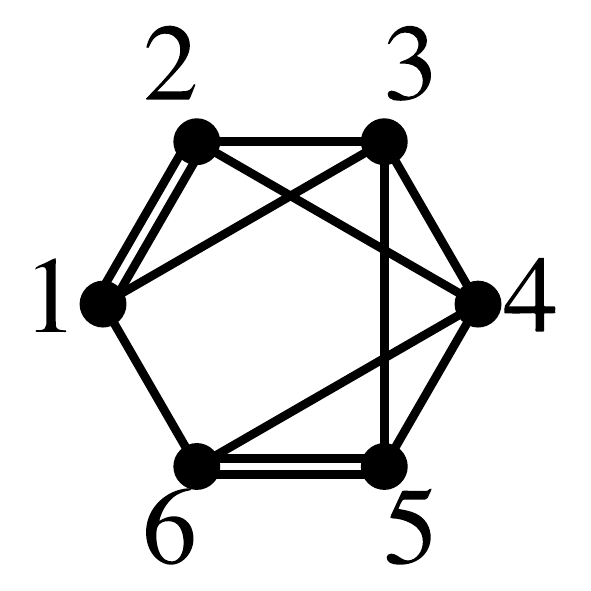}\!+\!\frac{s_{123}\!+\!s_{3(45)}}{s_{123}}\!\!\!\fig{-26.05pt}{0.35}{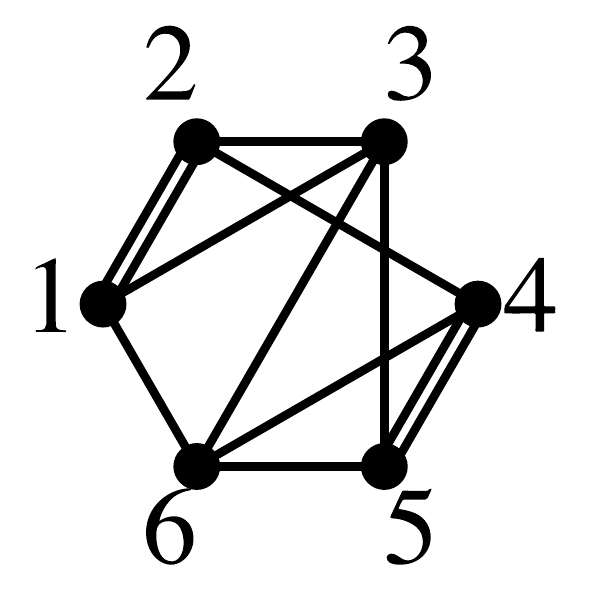}\!+\!\frac{s_{123}\!+\!s_{(23)4}}{s_{123}}\!\!\!\fig{-26.05pt}{0.35}{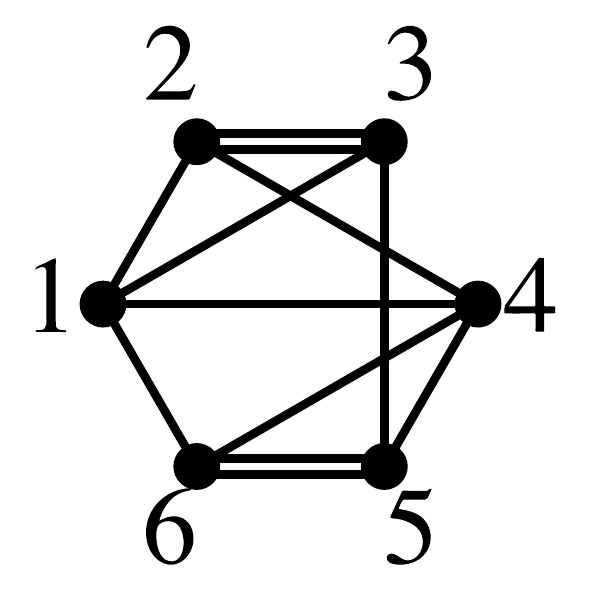}\nonumber\\
&\phantom{\,=\,}-\!\frac{s_{123}\!+\!s_{24}\!+\!s_{3(45)}}{s_{123}}\!\!\!\fig{-26.05pt}{0.35}{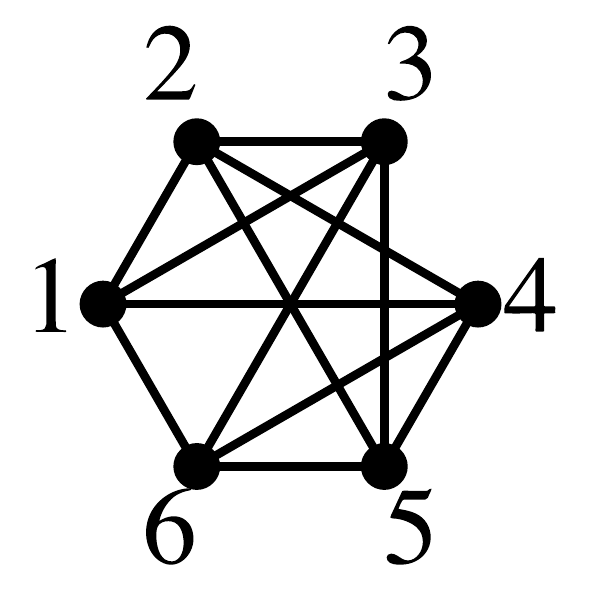}\!+\!\frac{s_{123}\!+\!s_{(23)(45)}}{s_{123}}\!\!\!\fig{-26.05pt}{0.35}{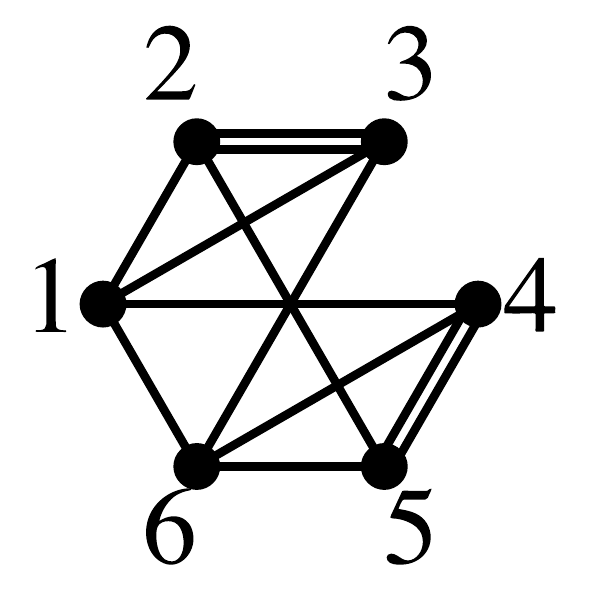}\\
&=-\frac{1}{s_{123}^2}\left(\frac{s_{123}\!+\!s_{34}}{s_{12}s_{56}}+\frac{s_{123}\!+\!s_{3(45)}}{s_{12}s_{45}}+\frac{s_{123}\!+\!s_{(23)4}}{s_{23}s_{56}}+\frac{s_{123}\!+\!s_{(23)(45)}}{s_{23}s_{45}}\right).\nonumber\end{align}
(Notice that the fourth term in the expansion above vanishes upon integration.) Similar reduction procedures exist for every integrand that we have checked---generating all terms necessary for amplitudes through 8 particles. For the sake of reference, we provide a complete analytic representation of the 6-particle amplitude in \mbox{Appendix \ref{appendix:6_pt_formula}}.

\section{Conclusions}\label{sec:conclusions}

In this paper, we have proposed a systematic algorithm to eliminate problematic $k$-tuples by integrand-level monodromy relations, which hold only at the support of scattering equations. Combining proper rewriting of diagonal entries of the $C$-matrix, we are able to write CHY integrand to a manifestly modular-invariant form and then using the integration rule given in \cite{Baadsgaard:2015voa,Baadsgaard:2015ifa,Baadsgaard:2015hia} to obtain an analytic CHY representation of Yang-Mills amplitudes. It is obvious that our method can be used in any theory, including gravity theories. 

One interesting aspect of this representation of Yang-Mills amplitudes is the following. Upon expanding the Pfaffian we get the sum of CHY integrands dressed with proper kinematic factors $s_{ij}$. Although potentially some CHY integrands could produce higher order poles ${1/s_A^2}$, the dressed kinematic factors conspire to reduce them to simple poles, as expected on physical grounds This is similar to the phenomenon observed in the KLT relations $A_L {\cal S} A_R$, where the momentum kernel ${\cal S}$ removes precisely removes double pole properly. Indeed, the momentum kernel~\cite{BjerrumBohr:2010hn} is directly related to the generator of BCJ-type identities~\cite{BjerrumBohr:2010ta}. It could be useful to understand the detailed mechanism in terms of CHY integrands further. 

Another intriguing direction is following. With our algorithm, it is straightforward to write down analytic expression for essentially any given CHY integrand. Thus, it maybe possible to consider a more general investigation of the mapping between the CHY formalism and general quantum field theories. Turning trees into loops, one can now also very explicitly consider loop amplitudes in this framework.

\section*{Acknowledgements}
The authors gratefully acknowledge helpful discussions with Christian Baadsgaard and David McGady. This work was supported in part by MOBILEX research grant from the Danish Council for Independent Research (JLB), but the National Science Foundation under Grand No.\ NSF PHY11-25915 (PHD), and by Qiu-Shi and Chinese NSF funding under contract Nos.\ {11575156, 11135006, and 11125523} (BF).

\newpage
\appendix\section{Explicit Representations of Yang-Mills Amplitudes}\label{appendix:explicit_formulae}
\subsection{Analytic CHY Representation of the Five-Particle Amplitude}\label{appendix:5_pt_formula}
Directly expanding (the manifestly-M\"{o}bius invariant form of) the CHY representation of the five-particle amplitude in Yang-Mills gives a total of 26 distinct integrands. Applying the rules described in this note and collecting terms into cyclic classes gives the following analytic representation for the amplitude,
\eq{\mathcal{A}_5\equiv\alpha_1\,\e{1}{2}\e{3}{4}+\alpha_2\,\e{1}{2}\e{3}{5}+\alpha_3\,\e13\e24+\beta_1\,\e{1}{2}+\beta_2\,\e{1}{3}+\text{{\it distinct} cyclic},}
where the coefficients are as follows:
\begin{equation}\begin{split}
\fwboxR{5pt}{\alpha_1\equiv}&\fwboxL{355pt}{\phantom{\pl}\ppfrac{\ek{5}{2} \ls{4}{1}{5} \pl \ek{5}{4} \s{2}{3} \mi \ek{5}{3} \s{2}{4}}{\s{1}{5} \s{3}{4}}\pl\ppfrac{\ek{5}{4} \s{2}{3}\mi\ek{5}{3} \s{2}{4} \pl \ek{5}{2} \s{4}{5}}{\s{1}{2} \s{3}{4}}}\\&\pl\ppfrac{\ek{5}{4} \s{2}{3}}{\s{1}{2} \s{4}{5}}\pl\ppfrac{\ek{5}{4}}{\s{4}{5}}\pl\ppfrac{\lek{5}{2}{4}}{\s{1}{5}}\pl\ppfrac{\ek{5}{2}}{\s{1}{2}}\,,
\end{split}\end{equation}
\begin{equation}\begin{split}
\fwboxR{5pt}{\alpha_2\equiv}&\fwboxL{355pt}{\phantom{\pl}\ppfrac{\lek{4}{2}{5} \ls{2}{4}{5} \mi \ek{4}{2} \s{1}{3}}{\s{1}{2} \s{4}{5}}\mi\ppfrac{\ek{4}{3} \s{2}{5}}{\s{1}{2} \s{3}{4}}\,,}
\end{split}\end{equation}
\begin{equation}\begin{split}
\fwboxR{5pt}{\alpha_3\equiv}&\fwboxL{355pt}{\phantom{\pl}\ppfrac{\ek{5}{1}}{\s{1}{5}}\mi\ppfrac{\ek{5}{4}}{\s{4}{5}}\,,}
\end{split}\end{equation}
\begin{equation}\begin{split}
\fwboxR{5pt}{\beta_1\equiv}&\fwboxL{355pt}{\phantom{\pl}\ppfrac{\ek{3}{2} \big[\ek{4}{1} \ek{5}{4}\mi\ek{4}{5} \ek{5}{1} \big]\pl \ek{3}{1} \big[\ek{4}{5} \ek{5}{2} \mi \ek{4}{2} \ek{5}{4}\big]}{\s{1}{2} \s{4}{5}}}\\&\pl\ppfrac{\ek{3}{2} \big[\ek{4}{1} \ek{5}{4}\mi\ek{4}{5} \ek{5}{1}\big]}{\s{2}{3} \s{4}{5}}\pl\ppfrac{\big[\ek{3}{2} \ek{4}{3}\mi\ek{3}{4} \ek{4}{2}\big] \ek{5}{1}}{\s{1}{5} \s{3}{4}}\\&\pl\ppfrac{\ek{4}{3} \big[\ek{3}{2} \ek{5}{1} \mi \ek{3}{1} \ek{5}{2}\big] \pl \ek{3}{4} \big[\ek{4}{1} \ek{5}{2}\mi\ek{4}{2} \ek{5}{1}\big]}{\s{1}{2} \s{3}{4}}\\&\mi\ppfrac{\ek{3}{2} \ek{5}{1} \lek{4}{1}{5}}{\s{1}{5} \s{2}{3}}\,,
\end{split}\end{equation}
\begin{equation}\begin{split}
\fwboxR{5pt}{\beta_2\equiv}&\fwboxL{355pt}{\phantom{\pl}\ppfrac{\ek{2}{1} \big[\ek{4}{5} \ek{5}{3} \mi \ek{4}{3} \ek{5}{4}\big]}{\s{1}{2} \s{4}{5}}\pl\ppfrac{\ek{2}{3} \big[\ek{4}{5} \ek{5}{1} \mi \ek{4}{1} \ek{5}{4}\big]}{\s{2}{3} \s{4}{5}}}\\&\pl\ppfrac{\ek{2}{3} \ek{5}{1} \lek{4}{1}{5}}{\s{1}{5} \s{2}{3}}\pl\ppfrac{\ek{4}{3} \ek{5}{1} \lek{2}{1}{5}}{\s{1}{5} \s{3}{4}}\mi\ppfrac{\ek{2}{1} \ek{4}{3} \lek{5}{3}{4}}{\s{1}{2} \s{3}{4}}\,.
\end{split}\end{equation}
We have verified this expression matches known results (e.g.~\cite{Medina:2002nk}, and BCFW~\cite{Bourjaily:2010wh}). Explicit, machine-readable expressions can be found in the {\sc Mathematica} notebook {\tt amplitude\_cyclic\_seeds.nb} included as part of this work's submission files to the {\tt arXiv}.

\newpage
\subsection{Analytic CHY Representation of the Six-Particle Amplitude}\label{appendix:6_pt_formula}
Directly expanding (the manifestly-M\"{o}bius invariant form of) the CHY representation of the six-particle amplitude in Yang-Mills gives a total of 237 distinct integrands. Applying the rules described in this note and collecting terms into cyclic classes gives the following analytic representation for the amplitude,\vspace{-2.5pt}
\begin{equation}\fwbox{0pt}{\begin{split}
\mathcal{A}_6\equiv&\phantom{+}\,\,\alpha_1\,\e{1}{2}\e{3}{4}\e56+\alpha_2\,\e{1}{2}\e{3}{5}\e46+\alpha_3\,\e12\e36\e45+\alpha_4\,\e13\e25\e46\\
&+\beta_1\,\e{1}{2}\e35+\beta_2\,\e14\e25+\beta_3\,\e13\e24+\beta_4\,\e{1}{2}\e34+\beta_5\,\e12\e45+\beta_6\,\e12\e46\\
&+\beta_7\,\e13\e46+\beta_8\,\e13\e46+\beta_9\,\e12\e36+\gamma_1\,\e12+\gamma_2\,\e13+\gamma_3\,\e14+\text{{\it distinct} cyclic},\\[-5pt]
\end{split}}\vspace{0pt}\end{equation}
where the coefficients are as follows:\footnote{Here we have introduced a notation $\lek{i}{j\cdots}{k}\!\equiv\!2\epsilon_i\!\cdot\!(k_j{+}\ldots{+}k_l)$. Explicit expressions can be found in the {\sc Mathematica} notebook {\tt amplitude\_cyclic\_seeds.nb} included as part of this work's submission files to the {\tt arXiv}.}
\vspace{-2.5pt}\begin{equation}\begin{split}
\fwboxR{5pt}{\alpha_1\equiv}&\fwboxL{367pt}{\phantom{\pl}\ppfrac{1}{\pp{1}{2}{3}}
\pl\ppfrac{1}{\s{1}{6}}
\pl\ppfrac{\ls{2}{3}{5}}{\pp{1}{2}{6} \s{1}{2}}
\pl\ppfrac{\ls{2}{3}{5}}{\pp{1}{2}{6} \s{1}{6}}
\pl\ppfrac{\ls{5}{2}{4}}{\pp{1}{5}{6} \s{1}{6}}
\pl\ppfrac{\ls{5}{2}{4}}{\pp{1}{5}{6} \s{5}{6}}
\pl\ppfrac{\s{2}{5}}{\s{1}{2} \s{5}{6}}
\pl\ppfrac{\s{4}{5}}{\pp{1}{2}{3} \s{5}{6}}
\pl\ppfrac{\s{2}{3}}{\pp{1}{2}{3} \s{1}{2}}}\\&
\pl\ppfrac{\s{2}{3} \s{4}{5}\mi\ls{4}{2}{3} \s{2}{5} \mi \s{2}{4} \s{3}{5}}{\pp{1}{5}{6} \s{1}{6} \s{3}{4}}
\pl\ppfrac{\s{2}{3} \s{4}{5}\mi\ls{4}{2}{3} \s{2}{5} \mi \s{2}{4} \s{3}{5} }{\pp{1}{5}{6} \s{3}{4} \s{5}{6}}\pl\ppfrac{\s{2}{3} \s{4}{5}}{\pp{1}{2}{3} \s{1}{2} \s{5}{6}}\\&
\pl\ppfrac{\ls{4}{5}{6} \s{2}{5} \mi \s{2}{4} \s{3}{5} \pl \s{2}{3} \s{4}{5}}{\s{1}{2} \s{3}{4} \s{5}{6}}
\pl\ppfrac{ \ls{2}{3}{5} \s{4}{5}\mi\s{2}{4} \s{3}{5}}{\pp{1}{2}{6} \s{1}{2} \s{3}{4}}
\pl\ppfrac{\ls{2}{3}{5} \s{4}{5}\mi\s{2}{4} \s{3}{5}}{\pp{1}{2}{6} \s{1}{6} \s{3}{4}}\,,\\[-8pt]
\end{split}\end{equation}
\vspace{-2.5pt}\begin{equation}\begin{split}
\fwboxR{5pt}{\alpha_2\equiv}&\fwboxL{367pt}{\phantom{\pl}\ppfrac{1}{\pp{1}{2}{6}}\mi\ppfrac{1}{\pp{1}{2}{3}}\mi\ppfrac{1}{\s{1}{6}}
\pl\ppfrac{\ls{2}{1}{6}}{\pp{1}{2}{6} \s{1}{6}}
\mi\ppfrac{\s{2}{3}}{\pp{1}{2}{3} \s{1}{2}}
\pl\ppfrac{\s{2}{6}}{\pp{1}{2}{6} \s{1}{2}}\,,}\\[-8pt]
\end{split}\end{equation}
\vspace{-2.5pt}\begin{equation}\begin{split}
\fwboxR{5pt}{\alpha_3\equiv}&\fwboxL{360pt}{\phantom{\pl}\ppfrac{1}{\pp{1}{2}{6}}
\pl\ppfrac{1}{\s{5}{6}}
\pl\ppfrac{\s{3}{4}}{\pp{1}{2}{6} \s{4}{5}}
\mi\ppfrac{\ls{2}{1}{6}}{\pp{1}{2}{6} \s{1}{2}}
\mi\ppfrac{\ls{2}{1}{6} \s{3}{4}}{\pp{1}{2}{6} \s{1}{2} \s{4}{5}}
\pl\ppfrac{\ls{2}{3}{4}}{\s{1}{2} \s{5}{6}}
\pl\ppfrac{\ls{4}{1}{3}}{\pp{1}{2}{3} \s{4}{5}}
\pl\ppfrac{\ls{4}{1}{3}}{\pp{1}{2}{3} \s{5}{6}}}\\&
\pl\ppfrac{\ls{4}{1}{3} \s{2}{3} \mi \s{1}{3} \s{2}{4}}{\pp{1}{2}{3} \s{1}{2} \s{4}{5}}
\pl\ppfrac{\ls{4}{1}{3} \s{2}{3} \mi \s{1}{3} \s{2}{4}}{\pp{1}{2}{3} \s{1}{2} \s{5}{6}}\,,\\[-8pt]
\end{split}\end{equation}
\begin{equation}\begin{split}
\fwboxR{5pt}{\alpha_4\equiv}&\fwboxL{360pt}{\phantom{\pl}\ppfrac{1}{\pp{1}{2}{3}}\,,}
\end{split}\end{equation}
\begin{equation}\begin{split}
\fwboxR{5pt}{\beta_1\equiv}&\fwboxL{360pt}{\phantom{\pl}\ppfrac{\ek{4}{1} \ek{6}{2}
\mi\ek{4}{2} \lek{6}{1}{3}}{\s{1}{2} \s{5}{6}}
\mi\ppfrac{\ek{4}{2} \ek{6}{1}
\mi \ek{4}{1} \ek{6}{2}}{\pp{1}{2}{6} \s{1}{2}}
\mi\ppfrac{\ek{4}{2} \ek{6}{1}}{\pp{1}{2}{6} \s{1}{6}}
\mi\ppfrac{\ek{4}{5} \lek{6}{1}{3}}{\pp{1}{2}{3} \s{4}{5}}}\\&\fwboxL{360pt}{
\pl\ppfrac{\ek{4}{2} \lek{6}{1}{3} \mi\lek{4}{1}{3} \lek{6}{2}{5}}{\pp{1}{2}{3} \s{5}{6}}
\pl\ppfrac{\ek{4}{3} \ek{6}{1} \ls{2}{3}{4}}{\pp{1}{2}{6} \s{1}{6} \s{3}{4}}
\pl\ppfrac{\ek{4}{3} \ek{6}{1} \ls{2}{3}{4}}{\pp{1}{5}{6} \s{1}{6} \s{3}{4}}
\mi\ppfrac{\ek{4}{5} \ek{6}{1}}{\s{1}{6} \s{4}{5}}}\\&\fwboxL{360pt}{
\pl\ppfrac{\ek{4}{3} \ek{6}{1} \ls{2}{3}{4}}{\pp{1}{5}{6} \s{3}{4} \s{5}{6}}
\mi\ppfrac{\ek{4}{3} \ek{6}{1}}{\pp{1}{2}{6} \s{3}{4}}
\pl\ppfrac{\ek{4}{3} \ek{6}{1}}{\pp{1}{5}{6} \s{1}{6}}
\pl\ppfrac{\ek{4}{3} \ek{6}{1}}{\pp{1}{5}{6} \s{5}{6}}
\mi\ppfrac{\ek{4}{3} \ek{6}{1}}{\s{3}{4} \s{5}{6}}}\\&\fwboxL{360pt}{
\pl\ppfrac{\ek{4}{3} \big[\lek{6}{2}{5} \s{2}{6}\mi\ek{6}{2} \ls{1}{3}{4}\mi \ek{6}{1} \s{2}{5} \big]}{\pp{1}{2}{6} \s{1}{2} \s{3}{4}}
\pl\ppfrac{\ek{4}{3} \big[\lek{6}{2}{5} \ls{2}{5}{6}\mi\ek{6}{2} \ls{1}{3}{4}\big]}{\s{1}{2} \s{3}{4} \s{5}{6}}}\\&\fwboxL{360pt}{
\pl\ppfrac{\ek{4}{5} \big[\ek{6}{2} \s{1}{3} \mi \ek{6}{1} \s{2}{3} \pl \ek{6}{3} \s{2}{6}\big]}{\pp{1}{2}{6} \s{1}{2} \s{4}{5}}
\pl\ppfrac{\ek{4}{5} \big[\ek{6}{2} \s{1}{3} \mi \lek{6}{1}{3} \s{2}{3}\big]}{\pp{1}{2}{3} \s{1}{2} \s{4}{5}}}\\&\fwboxL{360pt}{
\mi\ppfrac{\ek{4}{5} \ek{6}{1} \s{2}{3}}{\pp{1}{2}{6} \s{1}{6} \s{4}{5}}
\pl\ppfrac{\big[\ek{4}{2} \lek{6}{1}{3}\mi\ek{6}{2} \lek{4}{1}{3}\big] \ls{3}{1}{2} \pl \ek{6}{5} \big[\ek{4}{2} \s{1}{3} \mi \lek{4}{1}{3} \s{2}{3}\big]}{\pp{1}{2}{3} \s{1}{2} \s{5}{6}}\,,}
\end{split}\end{equation}
\begin{equation}\begin{split}
\fwboxR{5pt}{\beta_2\equiv}&\fwboxL{360pt}{\phantom{\pl}\ppfrac{\big[\ek{3}{2} \mi \ek{3}{4}\big] \ek{6}{5}}{\pp{1}{5}{6} \s{1}{6}}
\pl\ppfrac{\big[\ek{3}{2} \mi \ek{3}{4}\big] \ek{6}{5}}{\pp{1}{5}{6} \s{5}{6}}
\pl\ppfrac{\ek{3}{2} \ek{6}{5} \ls{4}{2}{3}}{\pp{1}{5}{6} \s{1}{6} \s{2}{3}}}\\&
\pl\ppfrac{\ek{3}{2} \ek{6}{5} \ls{4}{2}{3}}{\pp{1}{5}{6} \s{2}{3} \s{5}{6}}
\mi\ppfrac{\ek{3}{2} \lek{6}{1}{5}}{\s{1}{6} \s{2}{3}}
\mi\ppfrac{\ek{3}{4} \ek{6}{2} \ls{5}{3}{4}}{\pp{1}{2}{6} \s{1}{6} \s{3}{4}}
\mi\ppfrac{\ek{3}{4} \ek{6}{2}}{\pp{1}{2}{6} \s{1}{6}}\\&
\mi\ppfrac{\ek{3}{4} \ek{6}{5} \ls{2}{3}{4}}{\pp{1}{5}{6} \s{1}{6} \s{3}{4}}
\mi\ppfrac{\ek{3}{4} \ek{6}{5} \ls{2}{3}{4}}{\pp{1}{5}{6} \s{3}{4} \s{5}{6}}
\mi\ppfrac{\ek{3}{4} \lek{6}{3}{4}}{\s{1}{6} \s{3}{4}}\,,
\end{split}\end{equation}
\begin{equation}\begin{split}
\fwboxR{5pt}{\beta_3\equiv}&\fwboxL{360pt}{\phantom{\pl}\ppfrac{\ek{5}{4} \lek{6}{4}{5}}{\pp{1}{2}{3} \s{4}{5}}
\mi\ppfrac{\ek{5}{4} \ek{6}{1}}{\s{1}{6} \s{4}{5}}
\pl\ppfrac{\ek{5}{6} \ek{6}{1} \mi \ek{5}{1} \ek{6}{5}}{\pp{1}{5}{6} \s{5}{6}}}\\&
\pl\ppfrac{\ek{5}{4} \ek{6}{5}\mi\ek{5}{6} \ek{6}{4} }{\pp{1}{2}{3} \s{5}{6}}
\pl\ppfrac{\ek{6}{1} \lek{5}{1}{6}}{\pp{1}{5}{6} \s{1}{6}}\,,
\end{split}\end{equation}
\begin{equation}\begin{split}
\fwboxR{5pt}{\beta_4\equiv}&\fwboxL{360pt}{\phantom{\pl}\ppfrac{\big[\ek{5}{2} \ek{6}{1} \mi \ek{5}{1} \ek{6}{2}\big]\big(\s{5}{6}+\pp{1}{2}{6}\big)}{\pp{1}{2}{6} \s{1}{2} \s{5}{6}}
\pl\ppfrac{\ek{6}{1} \lek{5}{2}{4} \mi \ek{5}{1} \lek{6}{2}{4}}{\pp{1}{5}{6} \s{5}{6}}}\\&\fwboxL{360pt}{
\pl\ppfrac{1}{\pp{1}{5}{6} \s{3}{4} \s{5}{6}}\Big[\big[\ek{5}{2} \ek{6}{1} \mi \ek{5}{1} \ek{6}{2}\big] \s{1}{4} \pl \big[\ek{5}{4} \ek{6}{1}
\mi \ek{5}{1} \ek{6}{4}\big] \s{2}{3}}\\&\fwboxL{360pt}{
\pl\big[\ek{5}{1} \ek{6}{3}\mi \ek{5}{3} \ek{6}{1}  \big] \s{2}{4} \pl \big[\ek{5}{2} \ek{6}{1} \mi \ek{5}{1} \ek{6}{2}\big] \s{4}{5}}\\&\fwboxL{360pt}{
\pl \big[\ek{5}{2} \ek{6}{1}\mi \ek{5}{1} \ek{6}{2}\big] \s{4}{6}\Big]
\pl\ppfrac{\ek{5}{2} \ek{6}{1}}{\pp{1}{2}{6} \s{1}{6}}
\pl\ppfrac{\ek{5}{4} \ek{6}{1}}{\s{1}{6} \s{4}{5}}
\pl\ppfrac{\ek{5}{4} \ek{6}{1} \s{2}{3}}{\pp{1}{2}{6} \s{1}{6} \s{4}{5}}}\\&\fwboxL{360pt}{
\pl\ppfrac{\ek{5}{4} \ek{6}{3}}{\pp{1}{2}{6} \s{4}{5}}
\pl\ppfrac{\big[\ek{5}{2} \ek{6}{4}\mi\ek{5}{4} \ek{6}{2} \big] \s{1}{3} \pl\big[ \ek{5}{4} \lek{6}{1}{3}\mi \ek{6}{4} \lek{5}{1}{3}\big] \s{2}{3}}{\pp{1}{2}{3} \s{1}{2} \s{5}{6}}}\\&\fwboxL{360pt}{
\pl\ppfrac{1}{\s{1}{2} \s{3}{4} \s{5}{6}}\Big[\big[\ek{5}{3} \ek{6}{2}\mi \ek{5}{2} \ek{6}{3} \big]\s{1}{4}
\pl \big[\ek{5}{4} \ek{6}{1} \pl \ek{5}{4}\ek{6}{3} \big]\s{2}{3}}\\&\fwboxL{360pt}{
\mi \ek{6}{4} \big[\ek{5}{3} \ls{2}{3}{4} \mi \ek{5}{2} \s{1}{3} \pl \ek{5}{1} \s{2}{3}\big]\pl \big[\ek{5}{2} \ek{6}{1} \mi \ek{5}{1} \ek{6}{2}\big] \s{4}{6}}\\&\fwboxL{360pt}{
\pl\big[\ek{5}{4} \ek{6}{3}\mi \ek{5}{3} \ek{6}{1}\pl \ek{5}{1} \ek{6}{3}\big] \s{2}{4} \pl \big[\ek{5}{2} \ek{6}{1} \mi \ek{5}{1} \ek{6}{2}\big] \s{4}{5}}\\&\fwboxL{360pt}{
\mi\ek{5}{4} \ek{6}{2} \s{1}{3}\Big]
\pl\ppfrac{\ek{5}{4} (\lek{6}{1}{3} \s{2}{3}\mi\ek{6}{2} \s{1}{3})}{\pp{1}{2}{3} \s{1}{2} \s{4}{5}}
\pl\ppfrac{\ek{5}{4} \ek{6}{3} \mi \ek{5}{3} \ek{6}{4}}{\pp{1}{2}{6} \s{3}{4}}}\\&\fwboxL{360pt}{
\pl\ppfrac{1}{\pp{1}{2}{6} \s{1}{2} \s{3}{4}}\Big[\ek{6}{1} \big[\ek{5}{4} \s{2}{3} \pl \ek{5}{2} \s{4}{5}\big]
\mi\ek{5}{4} \ek{6}{3} \ls{2}{1}{6} \mi \ek{5}{3} \ek{6}{4} \ls{2}{3}{5}}\\&\fwboxL{360pt}{
\mi \ek{5}{3} \lek{6}{1}{4} \s{2}{4}
\mi \ek{6}{2} \big[\ek{5}{4} \s{1}{3} \mi \ek{5}{3} \s{1}{4} \pl \ek{5}{1} \s{4}{5}\big]\Big]
\pl\ppfrac{\ek{5}{4} \lek{6}{1}{3}}{\pp{1}{2}{3} \s{4}{5}}}\\&\fwboxL{360pt}{
\pl\ppfrac{\ek{5}{4} \ek{6}{3} \mi \ek{5}{3} \ek{6}{4}}{\s{3}{4} \s{5}{6}}
\pl\ppfrac{\ek{5}{4} \big[\ek{6}{1} \s{2}{3}\mi\ek{6}{3} \ls{2}{1}{6} \mi \ek{6}{2} \s{1}{3}\big]}{\pp{1}{2}{6} \s{1}{2} \s{4}{5}}}\\&\fwboxL{360pt}{
\mi\ppfrac{\ek{6}{1} \big[\ek{5}{2} \ls{4}{2}{3} \mi \ek{5}{4} \s{2}{3} \pl \ek{5}{3} \s{2}{4}\big]}{\pp{1}{5}{6} \s{1}{6} \s{3}{4}}
\pl\ppfrac{ \ek{5}{4} \lek{6}{1}{3}\mi\ek{6}{4} \lek{5}{1}{3} }{\pp{1}{2}{3} \s{5}{6}}}\\&\fwboxL{360pt}{
\pl\ppfrac{\ek{6}{1} \big[\ek{5}{4} \s{2}{3} \mi \ek{5}{3} \s{2}{4} \pl \ek{5}{2} \s{4}{5}\big]}{\pp{1}{2}{6} \s{1}{6} \s{3}{4}}
\pl\ppfrac{\ek{6}{1} \lek{5}{2}{4}}{\pp{1}{5}{6} \s{1}{6}}\,,}
\end{split}\end{equation}
\begin{equation}\begin{split}
\fwboxR{5pt}{\beta_5\equiv}&\fwboxL{360pt}{\phantom{\pl}\ppfrac{2 \ek{3}{4} \ek{6}{2} \pl \ek{3}{2} \lek{6}{1}{4}}{\pp{1}{2}{6} \s{1}{6}}
\mi\ppfrac{\ek{3}{1} \ek{6}{2} \pl \ek{3}{2} \lek{6}{2}{3}}{\pp{1}{2}{3} \s{1}{2}}
\pl\ppfrac{\big[\ek{3}{4}\mi\ek{3}{2} \big]\ek{6}{5}}{\pp{1}{5}{6} \s{1}{6}}}\\&\fwboxL{360pt}{
\pl\ppfrac{\big[\ek{3}{4}\mi\ek{3}{2} \big] \ek{6}{5}}{\pp{1}{5}{6} \s{5}{6}}
\mi\ppfrac{\ek{3}{2} \big[\ek{6}{5} \ls{4}{2}{3} \mi \ek{6}{4} \ls{5}{2}{3} \pl \lek{6}{2}{3} \s{5}{6}\big]}{\pp{1}{2}{3} \s{2}{3} \s{4}{5}}}\\&\fwboxL{360pt}{
\mi\ppfrac{\ek{3}{2} \big[\ek{6}{5} \ls{4}{2}{3} \pl \lek{6}{2}{3} \ls{5}{1}{6} \mi \ek{6}{4} \ls{5}{2}{3}\big]}{\s{1}{6} \s{2}{3} \s{4}{5}}
\mi\ppfrac{\ek{3}{2} \ek{6}{5} \ls{4}{2}{3}}{\pp{1}{2}{3} \s{2}{3} \s{5}{6}}}\\&\fwboxL{360pt}{
\mi\ppfrac{\ek{3}{2} \ek{6}{5} \ls{4}{2}{3}}{\pp{1}{5}{6} \s{1}{6} \s{2}{3}}
\mi\ppfrac{\ek{3}{2} \ek{6}{5} \ls{4}{2}{3}}{\pp{1}{5}{6} \s{2}{3} \s{5}{6}}
\mi\ppfrac{\ek{3}{2} \ek{6}{5}}{\s{1}{2} \s{5}{6}}
\mi\ppfrac{\ek{3}{2} \lek{6}{2}{3}}{\pp{1}{2}{3} \s{2}{3}}}\\&\fwboxL{360pt}{
\mi\ppfrac{\ek{3}{4} \big[\ek{6}{2} \ls{5}{1}{2}\mi\ek{6}{5} \ls{2}{3}{4} \pl \lek{6}{3}{4} \s{2}{5}]}{\pp{1}{2}{6} \s{1}{2} \s{3}{4}}
\pl\ppfrac{\ek{3}{4} \ek{6}{5} \ls{2}{3}{4}}{\pp{1}{5}{6} \s{3}{4} \s{5}{6}}
\pl\ppfrac{\ek{3}{4} \ek{6}{5}}{\pp{1}{2}{3} \s{5}{6}}}\\&\fwboxL{360pt}{
\pl\ppfrac{\ek{3}{4} \big[\ek{6}{5} \ls{2}{3}{4} \pl \ek{6}{2} (\ls{5}{1}{6} \pl \ls{5}{3}{4}) \mi \lek{6}{3}{4} \s{2}{5}\big]}{\pp{1}{2}{6} \s{1}{6} \s{3}{4}}
\pl\ppfrac{\ek{3}{4} \ek{6}{5} \ls{2}{3}{4}}{\pp{1}{5}{6} \s{1}{6} \s{3}{4}}}\\&\fwboxL{360pt}{
\pl\ppfrac{1}{\pp{1}{2}{6} \s{1}{2} \s{4}{5}}\Big[\ek{3}{4} \ek{6}{2} \ls{5}{1}{2} \mi\ek{3}{5} \ek{6}{2} \ls{4}{3}{6}
\mi (\ek{3}{2} \ek{6}{3} \pl \ek{6}{2} \lek{3}{1}{2}) \ls{5}{1}{6}}\\&\fwboxL{360pt}{
\pl 2 \ek{3}{4} \ek{6}{2} \ls{5}{3}{6} \mi \ek{3}{5} \ek{6}{4} \s{2}{3} \pl \ek{3}{5} \ek{6}{3} \s{2}{4} \mi \ek{3}{1} \ek{6}{2} \s{2}{5}}\\&\fwboxL{360pt}{
\mi \ek{3}{4} \ek{6}{3} \s{2}{5} \mi \ek{6}{4} \lek{3}{4}{5} \s{2}{5} \mi \ek{3}{2} \lek{6}{2}{3} \s{2}{5}\pl \ek{3}{2} \ek{6}{4} \s{3}{5}}\\&\fwboxL{360pt}{
\pl \ek{6}{5} (\ek{3}{4} \s{2}{3} \pl \lek{3}{4}{5} \s{2}{4} \mi \ek{3}{2} \s{3}{4})\Big]
\pl\ppfrac{\ek{3}{5} \ek{6}{2}}{\pp{1}{2}{6} \s{4}{5}}
\pl\ppfrac{ \ek{3}{4} \ek{6}{5}\mi\ek{3}{5} \ek{6}{4} }{\pp{1}{2}{3} \s{4}{5}}}\\&\fwboxL{360pt}{
\pl\ppfrac{\ek{3}{4} \ek{6}{5}\mi\ek{3}{5} \ek{6}{4}  }{\s{1}{6} \s{4}{5}}
\mi\ppfrac{\ek{6}{5} \big[\ek{3}{2} \s{3}{4}\mi\ek{3}{4} \s{2}{3} \pl \lek{3}{1}{2} \s{2}{4}\big]}{\pp{1}{2}{3} \s{1}{2} \s{5}{6}}}\\&\fwboxL{360pt}{
\pl\ppfrac{1}{\pp{1}{2}{6} \s{1}{6} \s{4}{5}}\Big[\ek{3}{5} \ek{6}{3} \s{2}{4}\mi\ek{3}{5} \ek{6}{4} \ls{2}{3}{5} \mi \ek{3}{2} \lek{6}{2}{3} \ls{5}{1}{2} \mi \ek{3}{5} \ek{6}{2} \s{2}{3}}\\&\fwboxL{360pt}{
\mi \ek{3}{4} \lek{6}{3}{4} \s{2}{5}\mi \ek{3}{5} \ek{6}{2} \s{3}{4} \pl \ek{6}{5} (\ek{3}{4} \s{2}{3} \pl \lek{3}{4}{5} \s{2}{4} \mi \ek{3}{2} \s{3}{4})}\\&\fwboxL{360pt}{
\mi \ek{3}{5} \ek{6}{2} \s{3}{5} \pl \ek{3}{2} \ek{6}{4} \s{3}{5} \mi \ek{6}{2} \big[\lek{3}{4}{5} \ls{5}{2}{4}\pl \ek{3}{4} \ls{5}{4}{6}\big] \pl \ek{3}{4} \ek{6}{2} \s{5}{6}}\\&\fwboxL{360pt}{
\mi \ek{3}{2} \lek{6}{2}{3} \s{5}{6}\Big]
\mi\ppfrac{\ek{6}{2} \big[ \lek{3}{1}{5}\mi2 \ek{3}{4}\big] \mi \ek{3}{2} \lek{6}{1}{4}}{\pp{1}{2}{6} \s{1}{2}}
\mi\ppfrac{\ek{3}{2} \lek{6}{2}{3}}{\s{1}{6} \s{2}{3}}}\\&\fwboxL{360pt}{
\pl\ppfrac{1}{\pp{1}{2}{3} \s{1}{2} \s{4}{5}}\Big[\ek{6}{4} \big[\lek{3}{1}{2} \s{2}{5}\mi\ek{3}{5} \s{2}{3} \pl \ek{3}{2} \s{3}{5}\big]\mi \ek{3}{1} \ek{6}{2} \s{5}{6}}\\&\fwboxL{360pt}{
\mi \ek{3}{2} \lek{6}{2}{3} \s{5}{6}
\mi\ek{6}{5} \big[\ek{3}{2} \s{3}{4}\mi\ek{3}{4} \s{2}{3} \pl \lek{3}{1}{2} \s{2}{4}\big]\Big]
\pl\ppfrac{\ek{3}{4} \ek{6}{5} \ls{2}{3}{4}}{\s{1}{2} \s{3}{4} \s{5}{6}}\,,}
\end{split}\end{equation}
\begin{equation}\begin{split}
\fwboxR{5pt}{\beta_6\equiv}&\fwboxL{360pt}{\phantom{\pl}\ppfrac{\ek{3}{1} \ek{5}{2}\mi\ek{3}{2} \ek{5}{1}}{\pp{1}{2}{3} \s{1}{2}}
\pl\ppfrac{ \ek{3}{1} \ek{5}{2}\mi\ek{3}{2} \ek{5}{1} }{\s{1}{2} \s{3}{4}}
\mi\ppfrac{\ek{3}{2} \ek{5}{1}}{\pp{1}{2}{3} \s{2}{3}}
\mi\ppfrac{\ek{3}{2} \ek{5}{4} \s{1}{6}}{\pp{1}{2}{3} \s{2}{3} \s{4}{5}}}\\&\fwboxL{360pt}{
\mi\ppfrac{\ek{3}{2} \ek{5}{4}}{\s{2}{3} \s{4}{5}}
\pl\ppfrac{\ek{3}{2} \ek{5}{6} \ls{1}{5}{6}}{\pp{1}{2}{3} \s{2}{3} \s{5}{6}}
\pl\ppfrac{\ek{3}{2} \ek{5}{6} \ls{1}{5}{6}}{\pp{1}{5}{6} \s{2}{3} \s{5}{6}}
\pl\ppfrac{\ek{3}{2} \ek{5}{6} \ls{1}{5}{6}}{\pp{1}{5}{6} \s{3}{4} \s{5}{6}}
\pl\ppfrac{\ek{3}{2} \ek{5}{6}}{\pp{1}{5}{6} \s{2}{3}}}\\&\fwboxL{360pt}{
\pl\ppfrac{\ek{3}{2} \ek{5}{6}}{\pp{1}{5}{6} \s{3}{4}}
\pl\ppfrac{\ek{3}{5} \lek{5}{2}{6} \mi \lek{3}{2}{6} \lek{5}{3}{4}}{\pp{1}{2}{6} \s{3}{4}}
\pl\ppfrac{\ek{3}{6} \ek{5}{2}}{\s{1}{2} \s{3}{4}}
\mi\ppfrac{\ek{3}{6} \ek{5}{4}}{\pp{1}{2}{3} \s{4}{5}}}\\&\fwboxL{360pt}{
\mi\ppfrac{\ek{5}{4} \lek{3}{2}{6}}{\pp{1}{2}{6} \s{4}{5}}
\pl\ppfrac{\ek{5}{6} \big[\ek{3}{2} \ls{1}{5}{6} \mi \ek{3}{1} \ls{2}{5}{6} \pl \lek{3}{5}{6} \s{2}{3}\big]}{\pp{1}{2}{3} \s{1}{2} \s{5}{6}}}\\&\fwboxL{360pt}{
\pl\ppfrac{\ek{5}{4} \big[\ek{3}{1} \s{2}{6}\mi\ek{3}{2} \s{1}{6} \mi \ek{3}{6} \s{2}{3}\big]}{\pp{1}{2}{3} \s{1}{2} \s{4}{5}}
\pl\ppfrac{\ek{5}{4} \big[\ek{3}{6} \ls{2}{1}{6} \mi \ek{3}{2} \s{1}{6} \pl \ek{3}{1} \s{2}{6}\big]}{\pp{1}{2}{6} \s{1}{2} \s{4}{5}}}\\&\fwboxL{360pt}{
\pl\ppfrac{\ek{5}{6} \big[\ek{3}{2} \ls{1}{5}{6} \pl \lek{3}{5}{6} \ls{2}{3}{4} \mi \ek{3}{1} \ls{2}{5}{6}\big]}{\s{1}{2} \s{3}{4} \s{5}{6}}
\pl\ppfrac{\ek{5}{6} \lek{3}{5}{6}}{\pp{1}{2}{3} \s{5}{6}}
\pl\ppfrac{\ek{5}{6} \lek{3}{5}{6}}{\s{3}{4} \s{5}{6}}}\\&\fwboxL{360pt}{
\pl\ppfrac{\lek{5}{3}{4} \big[\ek{3}{6} \ls{2}{1}{6} \mi \ek{3}{2} \s{1}{6} \pl \ek{3}{1} \s{2}{6}\big] \mi \ek{3}{5} \big[\ek{5}{6} \ls{2}{1}{6} \mi \ek{5}{2} \s{1}{6} \pl \ek{5}{1} \s{2}{6}\big]}{\pp{1}{2}{6} \s{1}{2} \s{3}{4}}\,,}
\end{split}\end{equation}
\begin{equation}\begin{split}
\fwboxR{5pt}{\beta_7\equiv}&\fwboxL{360pt}{\phantom{\pl}\ppfrac{\ek{4}{3} \lek{6}{1}{5}}{\s{3}{4} \s{5}{6}}
\mi\ppfrac{\ek{4}{3} \ek{6}{1} \ls{2}{3}{4}}{\pp{1}{5}{6} \s{1}{6} \s{3}{4}}
\mi\ppfrac{\ek{4}{3} \ek{6}{1} \ls{2}{3}{4}}{\pp{1}{5}{6} \s{3}{4} \s{5}{6}}
\mi\ppfrac{\ek{4}{3} \ek{6}{1}}{\pp{1}{5}{6} \s{1}{6}}}\\&
\mi\ppfrac{\ek{4}{3} \ek{6}{1}}{\pp{1}{5}{6} \s{5}{6}}
\pl\ppfrac{\ek{4}{5} \ek{6}{1}}{\s{1}{6} \s{4}{5}}
\mi\ppfrac{\ek{4}{5} \lek{6}{4}{5}}{\pp{1}{2}{3} \s{4}{5}}
\mi\ppfrac{\ek{6}{5} \lek{4}{5}{6}}{\pp{1}{2}{3} \s{5}{6}}\,,
\end{split}\end{equation}
\begin{equation}\begin{split}
\fwboxR{5pt}{\beta_8\equiv}&\fwboxL{360pt}{\phantom{\pl}\ppfrac{\ek{2}{1} \ek{5}{3}}{\pp{1}{2}{3} \s{1}{2}}
\mi\ppfrac{\ek{2}{1} \ek{5}{4} \ls{3}{4}{5}}{\pp{1}{2}{3} \s{1}{2} \s{4}{5}}
\mi\ppfrac{\ek{2}{1} \ek{5}{4} \ls{3}{4}{5}}{\pp{1}{2}{6} \s{1}{2} \s{4}{5}}
\mi\ppfrac{\ek{2}{1} \ek{5}{4}}{\pp{1}{2}{6} \s{1}{2}}
\pl\ppfrac{\ek{2}{1} \ek{5}{6} \s{3}{4}}{\pp{1}{2}{3} \s{1}{2} \s{5}{6}}}\\&\fwboxL{355pt}{\mi\ppfrac{\ek{2}{3} \ek{5}{6}}{\pp{1}{5}{6} \s{5}{6}}
\mi\ppfrac{\ek{2}{3} \lek{5}{1}{6} \ls{4}{2}{3}}{\pp{1}{5}{6} \s{1}{6} \s{2}{3}}
\pl\ppfrac{\ek{2}{3} \lek{5}{1}{6}}{\pp{1}{2}{3} \s{2}{3}}
\mi\ppfrac{\ek{2}{3} \lek{5}{1}{6}}{\pp{1}{5}{6} \s{1}{6}}
\pl\ppfrac{\ek{2}{3} \lek{5}{1}{6}}{\s{1}{6} \s{2}{3}}}\\&\fwboxL{355pt}{
\pl\ppfrac{\ek{2}{4} \ek{5}{6}}{\pp{1}{2}{3} \s{5}{6}}
\mi\ppfrac{\ek{5}{4} \lek{2}{1}{6} \ls{3}{4}{5}}{\pp{1}{2}{6} \s{1}{6} \s{4}{5}}
\pl\ppfrac{\ek{5}{4} \lek{2}{1}{6}}{\pp{1}{2}{3} \s{4}{5}}
\mi\ppfrac{\ek{5}{4} \lek{2}{1}{6}}{\pp{1}{2}{6} \s{1}{6}}
\pl\ppfrac{\ek{5}{4} \lek{2}{1}{6}}{\s{1}{6} \s{4}{5}}}\\&\fwboxL{355pt}{
\pl\ppfrac{\ek{2}{1} \ek{5}{6}}{\s{1}{2} \s{5}{6}}
\pl\ppfrac{\ek{2}{3} \ek{5}{4} \s{1}{6}}{\pp{1}{2}{3} \s{2}{3} \s{4}{5}}
\pl\ppfrac{\ek{2}{3} \ek{5}{4}}{\s{2}{3} \s{4}{5}}
\mi\ppfrac{\ek{2}{3} \ek{5}{6} \ls{4}{2}{3}}{\pp{1}{2}{3} \s{2}{3} \s{5}{6}}
\mi\ppfrac{\ek{2}{3} \ek{5}{6} \ls{4}{2}{3}}{\pp{1}{5}{6} \s{2}{3} \s{5}{6}}\,,}
\end{split}\end{equation}
\begin{equation}\begin{split}
\fwboxR{5pt}{\beta_9\equiv}&\fwboxL{360pt}{\phantom{\pl}\ppfrac{\ek{4}{3} \ek{5}{2} \mi \ek{4}{2} \ek{5}{3}}{\pp{1}{2}{6} \s{4}{5}}
\mi\ppfrac{\ek{4}{2} \ek{5}{1}\pl \ek{4}{1} \ek{5}{2}}{\s{1}{2} \s{4}{5}}
\mi\ppfrac{\ek{4}{3} \ek{5}{6} \ls{2}{5}{6}}{\s{1}{2} \s{3}{4} \s{5}{6}}}\\&\fwboxL{360pt}{
\pl\ppfrac{\big[\ek{4}{3} \ek{5}{2} \mi \ek{4}{2} \ek{5}{3}\big] \s{1}{6} \pl \big[\ek{5}{3} \lek{4}{1}{6}  \mi \ek{4}{3} \lek{5}{1}{6}\big] \s{2}{6}}{\pp{1}{2}{6} \s{1}{2} \s{4}{5}}}\\&\fwboxL{360pt}{\pl\ppfrac{\ek{4}{3} \ek{5}{2}}{\pp{1}{2}{6} \s{3}{4}}
\pl\ppfrac{\ek{4}{3} \big[\ek{5}{2} \s{1}{6} \mi \lek{5}{1}{6} \s{2}{6}\big]}{\pp{1}{2}{6} \s{1}{2} \s{3}{4}}
\pl\ppfrac{\ek{5}{6} \big[\lek{4}{1}{3} \ls{2}{1}{3} \mi \ek{4}{2} \s{1}{3}\big]}{\pp{1}{2}{3} \s{1}{2} \s{5}{6}}}\\&
\pl\ppfrac{\big[\ek{5}{6} \lek{4}{1}{3} \mi \ek{4}{6} \lek{5}{1}{3}\big] \ls{2}{1}{3} \pl \big[\ek{4}{6} \ek{5}{2} \mi \ek{4}{2} \ek{5}{6}\big] \s{1}{3}}{\pp{1}{2}{3} \s{1}{2} \s{4}{5}}\,,
\end{split}\end{equation}
\begin{equation}\begin{split}
\fwboxR{5pt}{\gamma_1\equiv}&\fwboxL{360pt}{\phantom{\pl}\ppfrac{1}{\pp{1}{2}{3} \s{1}{2} \s{5}{6}}\bigg[\big[\ek{3}{2} \ek{4}{1} \mi \ek{3}{1} \ek{4}{2}\big] \big[\ek{5}{6} \ek{6}{4} \mi \ek{5}{4} \ek{6}{5}\big]}\\&\fwboxL{360pt}{\pl \Big[\ek{3}{2} \big[ \ek{5}{1} \ek{6}{5}\mi\ek{5}{6} \ek{6}{1} \big]\pl \ek{3}{1} \big[\ek{5}{6} \ek{6}{2} \mi \ek{5}{2} \ek{6}{5}\big]\Big] \lek{4}{5}{6}\bigg]}\\&\fwboxL{360pt}{
\pl\ppfrac{1}{\pp{1}{2}{3} \s{1}{2} \s{4}{5}}\Big[\ek{3}{1} \ek{4}{1} \ek{5}{4} \ek{6}{2} \mi\ek{3}{2} \ek{4}{2} \ek{5}{4} \ek{6}{1} \mi \ek{3}{2} \ek{4}{3} \ek{5}{4} \ek{6}{1} }\\&\mi \ek{3}{2} \ek{4}{5} \ek{5}{4} \ek{6}{1} \mi \ek{3}{2} \ek{4}{5} \ek{5}{6} \ek{6}{1}\pl \ek{3}{1} \ek{4}{3} \ek{5}{4} \ek{6}{2}\\[5pt]&
\pl \ek{3}{1} \ek{4}{5} \ek{5}{4} \ek{6}{2} \pl \ek{3}{1} \ek{4}{5} \ek{5}{6} \ek{6}{2} \pl \ek{3}{2} \ek{4}{5} \ek{5}{1} \ek{6}{4} \\[5pt]&
\mi \ek{3}{1} \ek{4}{5} \ek{5}{2} \ek{6}{4}\pl \ek{3}{2} \ek{4}{1} \ek{5}{4} \lek{6}{2}{3}\mi \ek{3}{1} \ek{4}{2} \ek{5}{4} \lek{6}{1}{3}\\&\fwboxL{360pt}{
\pl \ek{4}{5} \big[\ek{3}{2} \ek{5}{1} \mi \ek{3}{1} \ek{5}{2}\big] \ek{6}{5} \Big]
\pl\ppfrac{1}{\pp{1}{2}{3} \s{2}{3} \s{4}{5}}\Big[\ek{3}{2} \big[\ek{4}{1} \ek{5}{4} \lek{6}{2}{3}}\\&\fwboxL{360pt}{
\mi\ek{4}{2} \ek{5}{4} \ek{6}{1} \mi \ek{4}{3} \ek{5}{4} \ek{6}{1} \mi \ek{4}{5} \ek{5}{4} \ek{6}{1} \mi \ek{4}{5} \ek{5}{6} \ek{6}{1}}\\&\fwboxL{360pt}{
\pl \ek{4}{5} \ek{5}{1} \ek{6}{4} \pl \ek{4}{5} \ek{5}{1} \ek{6}{5}\big]\Big]
\pl\ppfrac{\ek{3}{2} \big[\ek{5}{6} \ek{6}{1} \mi \ek{5}{1} \ek{6}{5}\big] \lek{4}{2}{3}}{\pp{1}{5}{6} \s{2}{3} \s{5}{6}}}\\&\fwboxL{360pt}{
\pl\ppfrac{\ek{3}{2} \ek{6}{1} \big[ \ek{4}{5} \lek{5}{2}{3}\mi\ek{5}{4} \lek{4}{2}{3}\big]}{\s{1}{6} \s{2}{3} \s{4}{5}}
\pl\ppfrac{\big[\ek{3}{2} \ek{4}{3}\mi\ek{3}{4} \ek{4}{2}\big] \ek{6}{1} \lek{5}{1}{6}}{\pp{1}{5}{6} \s{1}{6} \s{3}{4}}}\\&\fwboxL{360pt}{
\pl\ppfrac{\ek{3}{2} \Big[ \ek{5}{6} \big[\ek{6}{1} \lek{4}{2}{3} \pl \ek{4}{1} \lek{6}{1}{4}\big]\mi\ek{6}{5} \big[\ek{5}{1} \lek{4}{2}{3} \pl \ek{4}{1} \lek{5}{1}{4}\big]\Big]}{\pp{1}{2}{3} \s{2}{3} \s{5}{6}}}\\&\fwboxL{360pt}{
\mi\ppfrac{\big[\ek{3}{4} \ek{4}{2} \mi \ek{3}{2} \ek{4}{3}\big] \big[\ek{5}{6} \ek{6}{1} \mi \ek{5}{1} \ek{6}{5}\big]}{\pp{1}{5}{6} \s{3}{4} \s{5}{6}}
\pl\ppfrac{\ek{3}{2} \ek{6}{1} \lek{4}{2}{3} \lek{5}{1}{6}}{\pp{1}{5}{6} \s{1}{6} \s{2}{3}}}\\&\fwboxL{360pt}{
\pl\ppfrac{1}{\s{1}{2} \s{3}{4} \s{5}{6}}\bigg[\ek{3}{4} \Big[\ek{4}{2} \big[\ek{4}{1} \big[\ek{5}{6} \ek{6}{2} \mi \ek{5}{2} \ek{6}{5}\big]\mi\ek{5}{6} \ek{6}{1} \pl \ek{5}{1} \ek{6}{5}\big]\Big]}\\&\fwboxL{360pt}{
\pl \ek{4}{3} \Big[\ek{3}{2} \big[\ek{5}{6} \ek{6}{1} \mi \ek{5}{1} \ek{6}{5}\big] \pl \ek{3}{1} \big[\mi\ek{5}{6} \ek{6}{2} \pl \ek{5}{2} \ek{6}{5}\big]\Big]\bigg]}\\&\fwboxL{360pt}{
\pl\ppfrac{1}{\pp{1}{2}{6} \s{1}{2} \s{3}{4}}\bigg[\big[\ek{3}{5} \ek{4}{3} \mi \ek{3}{4} \ek{4}{5}\big] \big[\ek{5}{2} \ek{6}{1} \mi \ek{5}{1} \ek{6}{2}\big]}\\&\fwboxL{360pt}{
\pl \Big[\ek{4}{3} \big[\ek{3}{1} \ek{6}{2}\mi\ek{3}{2} \ek{6}{1}\big] \pl \ek{3}{4} \big[\ek{4}{2} \ek{6}{1} \mi \ek{4}{1} \ek{6}{2}\big]\Big] \lek{5}{3}{4}\bigg]}\\&\fwboxL{360pt}{
\pl\ppfrac{1}{\pp{1}{2}{6} \s{1}{2} \s{4}{5}}\bigg[\big[\ek{4}{5} \ek{5}{3} \mi \ek{4}{3} \ek{5}{4}\big] \big[\ek{3}{2} \ek{6}{1} \mi \ek{3}{1} \ek{6}{2}\big]}\\&\fwboxL{360pt}{
\pl \Big[\ek{5}{4} \big[\ek{4}{2} \ek{6}{1} \mi \ek{4}{1} \ek{6}{2}\big] \pl \ek{4}{5} \big[\ek{5}{1} \ek{6}{2}\mi\ek{5}{2} \ek{6}{1} \big]\Big] \lek{3}{4}{5}\bigg]}\\&\fwboxL{360pt}{
\pl\ppfrac{\ek{6}{1} \Big[\ek{3}{2} \big[\ek{4}{5} \ek{5}{3} \mi \ek{4}{3} \ek{5}{4}\big] \pl \big[\ek{4}{2} \ek{5}{4}\mi\ek{4}{5} \ek{5}{2}\big] \lek{3}{4}{5}\Big]}{\pp{1}{2}{6} \s{1}{6} \s{4}{5}}}\\&\fwboxL{360pt}{
\pl\ppfrac{\ek{6}{1} \Big[\big[\ek{3}{5} \ek{4}{3} \mi \ek{3}{4} \ek{4}{5}\big] \ek{5}{2} \pl \big[\ek{3}{4} \ek{4}{2} \mi \ek{3}{2} \ek{4}{3}\big] \lek{5}{3}{4}\Big]}{\pp{1}{2}{6} \s{1}{6} \s{3}{4}}\,,}
\end{split}\end{equation}
\begin{equation}\begin{split}
\fwboxR{5pt}{\gamma_2\equiv}&\fwboxL{360pt}{\phantom{\pl}\ppfrac{\ek{2}{1} \Big[\ek{4}{3} \big[\ek{5}{4} \ek{6}{5}\mi\ek{5}{6} \ek{6}{4}  \big ]\pl \big[\ek{5}{6} \ek{6}{3} \mi \ek{5}{3} \ek{6}{5}\big] \lek{4}{5}{6}\Big]}{\pp{1}{2}{3} \s{1}{2} \s{5}{6}}}\\&
\fwboxL{360pt}{\pl\ppfrac{\ek{2}{1} \Big[\big[  \ek{4}{5} \ek{5}{6}\mi\ek{4}{6} \ek{5}{4}\big] \ek{6}{3} \pl \big[ \ek{4}{3} \ek{5}{4}\mi\ek{4}{5} \ek{5}{3} \big] \lek{6}{4}{5}\Big]}{\pp{1}{2}{3} \s{1}{2} \s{4}{5}}}\\&
\fwboxL{360pt}{\pl\ppfrac{\ek{2}{3} \Big[\ek{4}{1} \big[ \ek{5}{4} \ek{6}{5}\mi\ek{5}{6} \ek{6}{4} \big] \pl \big[\ek{5}{6} \ek{6}{1} \mi \ek{5}{1} \ek{6}{5}\big] \lek{4}{5}{6}\Big]}{\pp{1}{2}{3} \s{2}{3} \s{5}{6}}}\\&
\fwboxL{360pt}{\pl\ppfrac{\ek{2}{3} \Big[\big[ \ek{4}{5} \ek{5}{6}\mi\ek{4}{6} \ek{5}{4}\big] \ek{6}{1} \pl \big[ \ek{4}{1} \ek{5}{4}\mi\ek{4}{5} \ek{5}{1}\big]\lek{6}{4}{5}\Big]}{\pp{1}{2}{3} \s{2}{3} \s{4}{5}}}\\&
\mi\ppfrac{\ek{4}{3} \ek{6}{1} \lek{2}{1}{6} \lek{5}{3}{4}}{\pp{1}{2}{6} \s{1}{6} \s{3}{4}}
\pl\ppfrac{\ek{2}{1} \ek{4}{3} \big[\ek{6}{5} \lek{5}{3}{4} \mi \ek{5}{6} \lek{6}{3}{4}\big]}{\s{1}{2} \s{3}{4} \s{5}{6}}\\&
\mi\ppfrac{\ek{4}{3} \ek{6}{1} \lek{2}{3}{4} \lek{5}{1}{6}}{\pp{1}{5}{6} \s{1}{6} \s{3}{4}}
\pl\ppfrac{\ek{2}{1} \big[\ek{4}{5} \ek{5}{3}\mi \ek{4}{3} \ek{5}{4}\big] \lek{6}{1}{2}}{\pp{1}{2}{6} \s{1}{2} \s{4}{5}}\\&
\mi\ppfrac{\ek{2}{3} \ek{6}{1} \lek{4}{2}{3} \lek{5}{1}{6}}{\pp{1}{5}{6} \s{1}{6} \s{2}{3}}
\pl\ppfrac{\ek{2}{3} \big[\ek{5}{1} \ek{6}{5}\mi\ek{5}{6} \ek{6}{1}\big]\lek{4}{2}{3}}{\pp{1}{5}{6} \s{2}{3} \s{5}{6}}\\&
\mi\ppfrac{\ek{2}{1} \ek{4}{3} \lek{5}{3}{4} \lek{6}{1}{2}}{\pp{1}{2}{6} \s{1}{2} \s{3}{4}}
\pl\ppfrac{\ek{2}{3} \ek{6}{1} \big[ \ek{4}{5} \lek{5}{1}{6}\mi\ek{5}{4} \lek{4}{1}{6}\big]}{\s{1}{6} \s{2}{3} \s{4}{5}}\\&
\fwboxL{360pt}{\pl\ppfrac{\ek{4}{3} \big[ \ek{5}{1} \ek{6}{5}\mi\ek{5}{6} \ek{6}{1}\big] \lek{2}{3}{4}}{\pp{1}{5}{6} \s{3}{4} \s{5}{6}}
\pl\ppfrac{\big[\ek{4}{5} \ek{5}{3} \mi \ek{4}{3} \ek{5}{4}\big] \ek{6}{1} \lek{2}{1}{6}}{\pp{1}{2}{6} \s{1}{6} \s{4}{5}}\,,}
\end{split}\end{equation}
\begin{equation}\begin{split}
\fwboxR{5pt}{\gamma_3\equiv}&\fwboxL{360pt}{
\phantom{\pl}\ppfrac{\big[\ek{5}{4} \ek{6}{5}\mi\ek{5}{6} \ek{6}{4}\big]\ek{2}{1}  \lek{3}{1}{2}}{\pp{1}{2}{3} \s{1}{2} \s{5}{6}}
\pl\ppfrac{\big[ \ek{2}{1} \ek{3}{2}\mi\ek{2}{3} \ek{3}{1} \big] \ek{5}{4} \lek{6}{4}{5}}{\pp{1}{2}{3} \s{2}{3} \s{4}{5}}}\\&\fwboxL{360pt}{
\pl\ppfrac{\big[\ek{2}{3} \ek{3}{1} \mi \ek{2}{1} \ek{3}{2}\big] \big[\ek{5}{6} \ek{6}{4} \mi \ek{5}{4} \ek{6}{5}\big]}{\pp{1}{2}{3} \s{2}{3} \s{5}{6}}
\pl\ppfrac{\ek{3}{4} \ek{6}{1} \lek{2}{3}{4} \lek{5}{1}{6}}{\pp{1}{5}{6} \s{1}{6} \s{3}{4}}}\\&\fwboxL{360pt}{\pl\ppfrac{\ek{3}{4} \ek{6}{1} \lek{2}{1}{6} \lek{5}{3}{4}}{\pp{1}{2}{6} \s{1}{6} \s{3}{4}}
\pl\ppfrac{\ek{2}{1} \ek{3}{4} \lek{5}{3}{4} \lek{6}{1}{2}}{\pp{1}{2}{6} \s{1}{2} \s{3}{4}}
\pl\ppfrac{\ek{2}{1} \ek{5}{4} \lek{3}{1}{2} \lek{6}{4}{5}}{\pp{1}{2}{3} \s{1}{2} \s{4}{5}}}\\&
\pl\ppfrac{\ek{2}{1} \ek{3}{4}\big [  \ek{5}{6} \lek{6}{3}{4}\mi\ek{6}{5} \lek{5}{3}{4}\big]}{\s{1}{2} \s{3}{4} \s{5}{6}}
\pl\ppfrac{\ek{2}{1} \ek{5}{4} \lek{3}{4}{5} \lek{6}{1}{2}}{\pp{1}{2}{6} \s{1}{2} \s{4}{5}}\\&\fwboxL{360pt}{
\pl\ppfrac{\big[\ek{2}{3} \ek{3}{4}\mi\ek{2}{4} \ek{3}{2}\big]\big[\ek{5}{6} \ek{6}{1} \mi \ek{5}{1} \ek{6}{5}\big]}{\pp{1}{5}{6} \s{2}{3} \s{5}{6}}
\pl\ppfrac{\ek{5}{4} \ek{6}{1} \lek{2}{1}{6} \lek{3}{4}{5}}{\pp{1}{2}{6} \s{1}{6} \s{4}{5}}}\\&\fwboxL{360pt}{
\pl\ppfrac{\big[\ek{2}{3} \ek{3}{4}\mi\ek{2}{4} \ek{3}{2}\big] \ek{6}{1} \lek{5}{1}{6}}{\pp{1}{5}{6} \s{1}{6} \s{2}{3}}
\pl\ppfrac{\big[\ek{5}{6} \ek{6}{1} \mi \ek{5}{1} \ek{6}{5}\big] \ek{3}{4} \lek{2}{3}{4}}{\pp{1}{5}{6} \s{3}{4} \s{5}{6}}}\\&
\pl\ppfrac{\ek{5}{4} \ek{6}{1} \big[\ek{2}{3} \lek{3}{1}{6}\mi\ek{3}{2} \lek{2}{1}{6}\big]}{\s{1}{6} \s{2}{3} \s{4}{5}}\,.
\end{split}\end{equation}

\section{Elaboration of the Monodromy Relation}\label{appendix:monodromy_elaboration}

In this section, we provide some further details on the identity (\ref{general_2_tuple_relation}). 
After some simplification this identity becomes
\eq { 
0=\left( \s21 {\z13\over \z12
\z23}+\sum_{k=3}^{n-1}\left(\sum_{i=1}^{k} \s{2}{i}\right) { \z{k}{k+1}\over
\z{k}{2} \z{2}{k+1}} \right)~.~~~\label{temp-1}}
Collecting coefficients of each $s_{2i}$ ({\it e.g.} $\s21$) we arrive at 
\eq{ {\z13\over \z12 \z23}+ \sum_{k=3}^{n-1}{ \z{k}{k+1}\over
\z{k}{2} \z{2}{k+1}}={\z{1}{n}\over \z12 \z2{n}}\,,}
where we have used the  following identity
\eq{ \sum_{k=a}^{n-1}{ \z{k}{k+1}\over \z{k}{2} \z{2}{k+1}}={
\z{a}{n}\over \z{a}{2} \z{2}{n}}~.~~\label{useful-identity-1}}
Similarly, the coefficient of $\s{2}{j}$, $j=3,..., (n-1)$ is
${\z{j}{n}\over \z{j}{2} \z{2}{n}}$.  Inserting the identity (\ref{temp-1}) we then 
have
\eq{ 0= \s21{\z1{n}\over \z12 \z{2}{n}}+ \sum_{j=3}^{n-1} \s2{j} {
\z{j}{n}\over \z{j}{2} \z{2}{n}}~.~~\label{temp-2}}
To prove (\ref{temp-2}) we use the scattering equation $-{\s21\over \z21}=\sum_{j=3}^{n} {\s{2}{j}\over \z{2}{j}}$
as the follows\footnote{In fact, (\ref{temp-2}) can be written as an identity of cross ratio $0= \s21+ \sum_{j=3}^{n-1} \s{2}{j} {
\z{j}{n}\z12\over \z{j}{2} \z{1}{n}}$. Such an identity and its generalizations will be systematically studied in the forthcoming ref.~\cite{CFGH}.}
\eqs { & \sum_{j=3}^{n-1} \s{2}{j}\left({
\z{j}{n}\over \z{j}{2} \z{2}{n}}+ {\z{1}{n}\over \z{2}{j} \z{2}{n}}\right)
+\s{2}{n} {\z{1}{n}\over \z{2}{n} \z{2}{n}}\\&= \sum_{j=3}^{n-1} \s{2}{j}
{\z{1}{2}+\z{2}{j}\over \z{2}{j} \z{2}{n}} +\s{2}{n} {\z{1}{n}\over \z{2}{n}
\z{2}{n}}\nonumber \\
 &=  {1\over \z{2}{n}}\sum_{j=3}^{n-1} \s{2}{j}+{\z{1}{2}\over
\z{2}{n}}\sum_{j=3}^{n-1} {\s{2}{j}\over \z{2}{j}} +\s{2}{n} {\z{1}{n}\over
\z{2}{n} \z{2}{n}}=0\,. }
The more general monodromy relation (\ref{general_k_tuple_relation}) can also be proved in a 
similar fashion highlighting  the deep intimacy between  the monodromy relations and the scattering 
equations.

We also note that in a systematic approach for the problematic $k$-tuples, we insert identities 
(\ref{general_k_tuple_id}) for each $k$-tuple when there are multiple ones. 
However, to avoid reproduce problematic $k$-tuples, we need to make these identities
compatible. For example, in the integrand 
\eq{{1\over \z12^3\z34^3\z56^3
\z23\z45\z61}\,,\label{term1}} 
there are three problematic $2$-tuples $\{1,2\}$, $\{3,4\}$ and $\{5,6\}$,
thus we need to use three identities of the type \ref{general_2_tuple_relation}. As an 
example a proper combination of three identities is given by 
\eqs{   PT(1,2,3,4,5,6) \label{6p-BCJ}& 
 \;=\;  \left(\left( { \ls213 \over \s12}+{
\ls256\ls423\over \s12\s34}\right){
\ls546\over \s56}+ { \ls256\ls416\over
\s12\s34}\right)PT(1,3,2,5,4,6)\nonumber \\
&\hskip-3.2cm -\left(\left( { \ls213 \over \s12}+{
\ls256\ls423\over
\s12\s34}\right){\ls513\over
\s56}-{\s26\ls416\over \s12\s34}\right) PT(1,3,5,2,4,6)\nonumber \\
& \hskip-3.2cm - \left( { \ls256\s14\over \s12\s34}
{\ls526\over \s56}-{\s26\s14\over
\s12\s34}\right) PT(1,3,5,2,6,4)
+\left({ \ls256\s41\over
\s12\s34}{\s45\over \s56} \right)PT(1,3,2,6,4,5)\nonumber \\&
 \hskip-3.2cm  -\left( { \ls213 \over \s12}+{
\ls256\ls423\over
\s12\s34}\right){\s15\over \s56} PT(1,5,3,2,4,6)+\left({
\ls256\s14\over \s12\s34}{\ls514\over
\s56}\right) PT(1,5,3,2,6,4)\nonumber \\ &
 \hskip-3.2cm -\left({\s26\ls435\over \s12\s34}\right)PT(1,3,5,4,2,6)~. \nonumber}
It is seen that all problematic $2$-tuples are removed. 
The result for the integration \ref{term1} is $\sum_{i=1}^7 T_i $, where 
\eqs{ T_1 & \;\; -\left(\left( { \ls213 \over \s12}+{
\ls256\ls423\over \s12\s34}\right){
\ls546\over \s56}+ { \ls256\ls416\over
\s12\s34}\right) 
  \times \left( {1\over \s12 \s34 \s56}+{1\over \s12
\pp123 \s56}\right)\,,\\
 T_2 & \;=\; \left(\left( { \ls213 \over \s12}+{
\ls256\ls423\over
\s12\s34}\right){\ls513\over
\s56}-{\s26\ls416\over \s12\s34}\right)\times
\left({1\over \s12\s34 \s56}\right)\,,\nonumber \\
 T_3 & \;=\;  -\left( -{
\ls256\s14\over \s12\s34} {\ls526\over
\s56}+{\s26\s14\over \s12\s34}\right)\times \left({1\over
\s12\s34 \s56}+{1\over \s34 \pp134\s56}\right)\,,  \nonumber \\
T_4 & \;=\; -\left( { \ls256\s14\over \s12\s34}{\s45\over
\s56} \right)\times \left({1\over \s12\s34\s56}+{1\over \s12
\pp123\s56}\right)\,,\\
 T_5 & \;=\;  \left( { \ls213 \over \s12}+{
\ls256\ls423\over
\s12\s34}\right){\s15\over \s56} \times \left({1\over \s12
\s34 \s56}+{1\over \s34\pp156\s56}\right)\,,\nonumber \\
 T_6 & \;=\;  \left({ \ls256\s14\over
\s12\s34}{\ls514\over \s56} \right)\times \left({1\over s_{12}
s_{34} s_{56}}\right)\,,\nonumber \\
T_7 & \;=\;  \left({\s26\ls435\over \s12\s34}\right)\times
\left({1\over \s12\s34\s56}+{1\over \s12\pp126
\s34}\right)\,.}

\newpage
%
\providecommand{\href}[2]{#2}\begingroup\raggedright\endgroup

\end{document}